\documentclass[aps,prd,showpacs,superscriptaddress,amsmath,amssymb]{revtex4}

\usepackage{slashed}
\usepackage[dvips]{color}
\usepackage[dvips]{epsfig}
\usepackage{latexsym}
\usepackage{bm}
\usepackage{upgreek}
\usepackage{mathrsfs}
\usepackage{times}
\usepackage{amsthm}
\usepackage{amssymb}
\usepackage{epsfig}
\usepackage{graphicx}
\usepackage{amsmath}

\usepackage{color}

\definecolor{purple}{rgb}{0.8,0,0.6}

\newcommand{\revision}[1]{{#1}}

\begin{document}

\title{Scalar excitation with Leggett frequency in $^3$ He-B and the $125$ GeV Higgs particle in top quark condensation models as Pseudo - Goldstone bosons}

\author{G.E.~Volovik}
\affiliation{Low Temperature Laboratory, Aalto University,  P.O. Box 15100, FI-00076 Aalto, Finland;}
\affiliation{Landau Institute for Theoretical Physics RAS,
 Kosygina 2,
119334 Moscow, Russia}

\author{M.A.~Zubkov}
\affiliation{Institute for Theoretical and Experimental Physics, B. Cheremushkinskaya 25, Moscow, 117259, Russia;}
\affiliation{Moscow Institute of Physics and Technology, 9, Institutskii per., Dolgoprudny, Moscow Region, 141700, Russia}
\affiliation{Far Eastern Federal University,  School of Biomedicine, 690950 Vladivostok, Russia}
\affiliation{CNRS, Laboratoire de Math\'ematiques et Physique Th\'eorique, Universit\'e Fran\c{c}ois-Rabelais, F\'ed\'eration Denis Poisson - CNRS, Parc de Grandmont, Universit\'e de Tours, 37200 France}

\begin{abstract}
We consider the scenario, in which the light Higgs scalar boson appears as the
Pseudo - Goldstone boson.
We discuss examples both in condensed matter and in relativistic field
theory.
In  $^3$He-B the symmetry breaking gives rise to 4  Nambu-Goldstone modes
and 14 Higgs modes.
At lower energy one of the four NG modes becomes the Higgs boson with small
mass. This is the mode measured in experiments with the longitudinal NMR,
and the Higgs mass  corresponds to the Leggett frequency $M_{\rm H}=\hbar
\Omega_B$. The formation of the Higgs mass is the result of the violation of
the hidden spin-orbit symmetry at low energy. In this scenario the symmetry
breaking energy scale $\Delta$ (the gap in the fermionic spectrum) and the
Higgs mass scale $M_{\rm H}$ are highly separated: $M_{\rm H}\ll \Delta$.
On the particle physics side we consider the model inspired by the models of \cite{dobrescu,Yamawaki}. At high energies the $SU(3)$ symmetry is assumed that relates the left  - handed top and bottom quarks to the additional fermion $\chi_L$. This symmetry is softly broken at low energies. As a result the only CP - even Goldstone boson acquires a mass and may be considered as the candidate for the role of the $125$ GeV scalar
boson. We consider the condensation pattern different from the one typical for the top - seesaw models, where the condensate $\langle \bar{t}_L\chi_R \rangle$ is off - diagonal. In our case the condensates are mostly diagonal. Unlike \cite{dobrescu,Yamawaki} the explicit mass terms are absent and the soft breaking of $SU(3)$ symmetry is given solely by the four - fermion terms. This reveals the complete analogy with $^3$He, where there is no explicit mass term and the spin - orbit interaction has the form of the four - fermion interaction.  \end{abstract}



\maketitle

\newcommand{\br}{{\bf r}}
\newcommand{\bu}{{\bf \delta}}
\newcommand{\bk}{{\bf k}}
\newcommand{\bq}{{\bf q}}
\def\({\left(}
\def\){\right)}
\def\[{\left[}
\def\]{\right]}

\newcommand{\barray}{\begin{eqnarray}}
\newcommand{\earray}{\end{eqnarray}}
\newcommand{\nn}{\nonumber \\}
\newcommand{\nl}{& \nonumber \\ &}
\newcommand{\bnl}{\right .  \nonumber \\  \left .}
\newcommand{\dbnl}{\right .\right . & \nonumber \\ & \left .\left .}

\newcommand{\beq}{\begin{equation}}
\newcommand{\eeq}{\end{equation}}
\newcommand{\ba}{\begin{array}}
\newcommand{\ea}{\end{array}}
\newcommand{\bea}{\begin{eqnarray}}
\newcommand{\eea}{\end{eqnarray} }
\newcommand{\be}{\begin{eqnarray}}
\newcommand{\ee}{\end{eqnarray} }
\newcommand{\bal}{\begin{align}}
\newcommand{\eal}{\end{align}}
\newcommand{\bi}{\begin{itemize}}
\newcommand{\ei}{\end{itemize}}
\newcommand{\ben}{\begin{enumerate}}
\newcommand{\een}{\end{enumerate}}
\newcommand{\bc}{\begin{center}}
\newcommand{\ec}{\end{center}}
\newcommand{\bt}{\begin{table}}
\newcommand{\et}{\end{table}}
\newcommand{\btb}{\begin{tabular}}
\newcommand{\etb}{\end{tabular}}
\newcommand{\bvec}{\left ( \ba{c}}
\newcommand{\evec}{\ea \right )}

\newcommand\e{{e}}
\newcommand\eurA{\eur{A}}
\newcommand\scrA{\mathscr{A}}

\newcommand\eurB{\eur{B}}
\newcommand\scrB{\mathscr{B}}

\newcommand\eurV{\eur{V}}
\newcommand\scrV{\mathscr{V}}
\newcommand\scrW{\mathscr{W}}

\newcommand\eurD{\eur{D}}
\newcommand\eurJ{\eur{J}}
\newcommand\eurL{\eur{L}}
\newcommand\eurW{\eur{W}}

\newcommand\eubD{\eub{D}}
\newcommand\eubJ{\eub{J}}
\newcommand\eubL{\eub{L}}
\newcommand\eubW{\eub{W}}

\newcommand\bmupalpha{\bm\upalpha}
\newcommand\bmupbeta{\bm\upbeta}
\newcommand\bmuppsi{\bm\uppsi}
\newcommand\bmupphi{\bm\upphi}
\newcommand\bmuprho{\bm\uprho}
\newcommand\bmupxi{\bm\upxi}

\newcommand\calJ{\mathcal{J}}
\newcommand\calL{\mathcal{L}}

\newcommand{\notyet}[1]{{}}

\newcommand{\sgn}{\mathop{\rm sgn}}
\newcommand{\tr}{\mathop{\rm Tr}}
\newcommand{\rk}{\mathop{\rm rk}}
\newcommand{\rank}{\mathop{\rm rank}}
\newcommand{\corank}{\mathop{\rm corank}}
\newcommand{\range}{\mathop{\rm Range\,}}
\newcommand{\supp}{\mathop{\rm supp}}
\newcommand{\p}{\partial}
\renewcommand{\P}{\grave{\partial}}
\newcommand{\yDelta}{\grave{\Delta}}
\newcommand{\yD}{\grave{D}}
\newcommand{\yeurD}{\grave{\eur{D}}}
\newcommand{\yeubD}{\grave{\eub{D}}}
\newcommand{\at}[1]{\vert\sb{\sb{#1}}}
\newcommand{\At}[1]{\biggr\vert\sb{\sb{#1}}}
\newcommand{\vect}[1]{{\bold #1}}
\def\R{\mathbb{R}}
\newcommand{\C}{\mathbb{C}}
\def\hvar{{\hbar}}
\newcommand{\N}{\mathbb{N}}\newcommand{\Z}{\mathbb{Z}}
\newcommand{\Abs}[1]{\left\vert#1\right\vert}
\newcommand{\abs}[1]{\vert #1 \vert}
\newcommand{\Norm}[1]{\left\Vert #1 \right\Vert}
\newcommand{\norm}[1]{\Vert #1 \Vert}
\newcommand{\Const}{{C{\hskip -1.5pt}onst}\,}
\newcommand{\sothat}{{\rm ;}\ }
\newcommand{\Range}{\mathop{\rm Range}}
\newcommand{\ftc}[1]{$\blacktriangleright\!\!\blacktriangleright$\footnote{AC: #1}}

\section{Introduction}

Spontaneous symmetry breaking gives rise to collective modes of the order
parameter field  -- the Higgs field. The oscillations of the Higgs field include the Nambu-Goldstone (NG) modes
-- the gapless phase modes which in gauge theories become massive gauge bosons
due to the Anderson-Higgs
mechanism; and the gapped amplitude modes -- the Higgs bosons.
The Higgs amplitude modes have been recently observed in electrically charged condensed matter system, the $s$-wave superconductor
\cite{Matsunaga2013,Matsunaga2014} (see also review paper \cite{PekkerVarma2015}), while they have been for a long time theoretically
\cite{Vdovin1963,Maki1974,Nagai1975,Tewordt-Einzel1976}
and experimentally \cite{Avenel1980,Lee1988,Collett2012} investigated
in electrically neutral superfluid phases of
$^3$He.

In superfluid phases of $^3$He the Higgs field  contains 18 real components. This
provides the arena for simulation of many phenomena in particle physics, including the physics of
the NG and Higgs bosons. In particular, superfluid $^3$He-A violates the conventional counting rule for the number of NG modes. In $^3$He-A the number of NG modes exceeds
the  number of broken symmetry generators, but it obeys the more general Novikov rule  \cite{Novikov1982},
according to which the number of NG modes
coincides with the dimension
of the ``tangent space'' in the space of the order parameter, see  the review paper \cite{VolovikZubkov2014}
and references therein.

Another example of the influence of superfluid  $^3$He is the connection between the fermionic
and bosonic masses in the theories with composite Higgs, which has been first formulated by Nambu after consideration of the $^3$He-B collective modes
\cite{Nambu1985}. If the Nambu sum rule is applicable to Standard Model, one may predict the masses of
extra Higgs bosons \cite{VolovikZubkovHiggs,VolovikZubkov2014}.

Here we discuss one more phenomenon -- the appearance of the
light Higgs bosons (LHB) as the pseudo NG modes. The origin of this phenomenon in $^3$He
is the hierarchy of energy scales, which exists in superfluid
$^3$He. In particular, the spin-orbit interaction is several orders of magnitude smaller than the
characteristic energy scale responsible for the formation of vacuum Higgs field \cite{VollhardtWolfle1990}.
When this interaction is neglected, the symmetry group of the
physical laws is enhanced, and the broken symmetry scheme in $^3$He-B gives rise to 4 NG modes
and 14 Higgs amplitude modes. The spin-orbit interaction reduces the symmetry and transforms one of the NG modes to the
Higgs mode with small mass.  The mechanism of the formation of the mass of the Higgs
boson $\#$15 in $^3$He-B is analogous to the little Higgs scenario \cite{LittleHiggsReview2005}.
The similar mechanism could be responsible for the
relatively small mass of the observed $125$ GeV  scalar
boson.
We consider the LH bosons in superfluid $^3$He-B. The parametric excitation
of the LH modes has been recently reported, which corresponds to the decay of magnon to two
light Higgses  \cite{Zavjalov2014}. We also consider the LH modes in the recently discovered
\cite{Dmitriev2012} polar phase of $^3$He in the nematically ordered  aerogel.

The idea, that Higgs boson of the SM may be composed of fermions follows the analogy with the models of superconductivity and superfluidity. In 1979 it was suggested, that Higgs boson is composed of additional technifermions \cite{technifermions}. This theory contains an additional set of fermions that interact with the Technicolor (TC) gauge bosons. This interaction is attractive and, therefore, by analogy with BCS superconductor theory it may lead to the formation of fermionic condensate. The TC theory suffers from the problems related to fermion mass generation. Extended Technicolor (ETC) interactions \cite{ETC} do not pass precision Electroweak tests due to the flavor changing neutral currents and due to the contributions to the Electroweak polarization operators. The so-called walking technicolor \cite{Walking} improves the situation essentially, but the ability to generate top quark mass remains problematical.

The idea, that Higgs boson may be composed of known SM fermions was suggested even earlier than Technicolor (in 1977) by H.Terazawa and co - authors \cite{Terazawa}. In the top quark condensation scenario, the top quark represents  the dominant component of the composite Higgs boson due to its large mass compared to the other components \cite{Terazawa2}. In 1989 this construction was recovered in  \cite{top}. Later the top quark condensation scenario was developed in a number of papers \cite{top2}.  In the conventional top – quark condensation models the scale of the new dynamics was assumed to be at about $10^{15}$ GeV. Such models typically predict the Higgs boson mass about $2 m_t \sim 350$ GeV \cite{Terazawa2,top,top2}, and they are excluded by present experimental data. In those models the prediction of Higgs boson mass is the subject of the large renormalization group corrections \cite{top2} due to the running of coupling constants between the working scale $10^{15}$ GeV and the electroweak scale $100$ GeV. But this running is not able to explain the appearance of the Higgs boson mass around 125 GeV.

In addition to the TC and the top – quark condensation models, models were developed \cite{top3}  (topcolor, topcolor assisted Technicolor, etc) that contain the elements of both mentioned approaches. Other models were suggested, in which the Higgs boson appears as the Goldstone boson of the broken approximate symmetry \cite{PNG} (for the realization of this idea in Little Higgs Models see \cite{LittleHiggs}).

It seems reasonable to look for a conceptually new model, in which Higgs bosons are composed (possibly, partially) of known SM fermions. Such a model may avoid difficulties of the models of Technicolor and the conventional models of top – quark condensation if it will be based on the analogy with certain condensed matter systems, like the superfluid $^3$He, in which the condensates are more complicated, than in the Technicolor models and the conventional models of top – quark condensation. (The latter models are based on the analogy with the simplest s-wave superconductors.)

Recently the models were proposed, that in a certain sense realize this idea \cite{dobrescu,Yamawaki}. In these models the Pseudo - Goldstone boson - the candidate for the role of the $125$ GeV Higgs boson appears in the framework of top seesaw \cite{topseesaw}. In both these papers the additional fermion $\chi$ is present typical for the top - seesaw models.   It has the quantum numbers of $t_R$ but if the gauge interactions of the Standard Model are neglected, its left - handed component may be considered together with $b_L$ and $t_L$ as the component of the $SU(3)$ triplet. As a result the structure of condensates is indeed more complicated than in the s-wave superconductor or in the simplest models of top quark condensation and is, therefore, to a certain extent similar to that of $^3$He.  The original inter - fermion  interactions of \cite{dobrescu,Yamawaki} are $SU(3)$ - symmetric. This symmetry is broken spontaneously giving rise to several Nambu - Goldstone bosons. Then the authors of \cite{dobrescu,Yamawaki} introduce the terms that softly break the $SU(3)$ symmetry explicitly (in particular, the explicit mass term for $\chi$ is added). As a result, one of the Goldstone bosons acquires a mass that may be smaller than $2 m_t$. Such a state is considered as a candidate for the role of the $125$ GeV Higgs boson.

In the present paper we consider the model inspired by the models of \cite{dobrescu} and \cite{Yamawaki}. In our case the original $SU(3)$ symmetry is broken explicitly by the additional four - fermion interaction instead of the explicit mass terms. We investigate the resulting model in the leading order of $1/N_c$ expansion. It is shown that the CP - even pseudo - Goldstone boson may have mass equal to $125$ GeV while the branching ratios of its decays do not contradict the present LHC data. We consider the condensation pattern different from the one typical for the top - seesaw models with the off - diagonal condensate $\langle \bar{t}_L\chi_R \rangle$. In our case the condensates are mostly diagonal.

It is worth mentioning that the considered model is of the Nambu - Jona - Lasinio (NJL) type, that is it contains the  effective 4 - fermion interaction \cite{Nambu1989}. The use of the one - loop approximation may cause a confusion because formally the contributions of higher loops to various physical quantities are strong. In \cite{cveticreview,cvetic} it has been shown that the next to leading (NTL) order approximation to the  fermion mass $m_f$ is weak compared to the one - loop approximation only if this mass is of the order of the cutoff $m_f \sim \Lambda$. It follows from analytical results and from numerical simulations made within the lattice regularization \cite{latticeNJL} that the dimensional physical quantities in the relativistic NJL models are typically of the order of the cutoff unless their small values  are protected by symmetry.

In the model of the present paper formally the one - loop results cannot be used because the cutoff is assumed to be many orders of magnitude larger than the generated fermion mass. That means, that in order to use the one - loop results we should start from the action of the model with the additional counter - terms that cancel dangerous quadratic divergences in the next to leading orders of $1/N_c$ expansion. Then the one - loop results give reasonable estimates to the physical quantities. Such a redefined NJL model is equivalent to the original NJL model defined in zeta or dimensional regularization. The four fermion coupling constants of the two regularizations are related by the finite renormalization (see \cite{ZetaNeutrino}, Appendix, Sect. 4.2.).
The NJL models in zeta regularization were considered,  in \cite{ZetaNeutrino,ZetaHiggs}. The NJL model in dimensional regularization was considered, for example, in \cite{NJLdim}.

It is generally assumed that there is the exchange by massive gauge bosons behind the NJL models of top quark condensation, top seesaw, and ETC. The appearance of the one - loop gap equation of NJL model may follow from the direct investigation of the theory with massive gauge fields interacting with fermions. Indeed, recently the indications were found that in the theory with exchange by massive gauge bosons the NJL approximation may be applied understood through its one - loop expressions \cite{Z2014_MGB}. Anyway, we assume that the model with the four - fermion interactions considered here should be explored in this way, i.e. the higher orders in $1/N_C$ contributions are simply disregarded. We suppose, that such an effective model appears as an approximation to a certain unknown renormalizable microscopic theory. For the further discussion of this issue see \cite{VolovikZubkov2014,VolovikZubkovHiggs,ZetaHiggs} and references therein.

The paper is organized as follows. In Section \ref{secthe} we discuss the appearance of the pseudo - Goldstone boson in superfluid phases of $^3$He due to the spin - orbit interaction. In Section
\ref{seesaw} we consider the model, in which the Pseudo - Goldstone boson composed  of top quark and the heavy fermion $\chi$ plays the role of the $125$ GeV Higgs boson.  In Section \ref{sectconclusions} we end with the conclusions.

\section{Superfluid  $^3$He}
\label{secthe}
\subsection{"Hydrodynamic action" in $^3$He (neglected spin-orbit interaction).}

According to \cite{He3} Helium - 3 may be described by the effective theory with the action
\begin{equation}
S_{} = \sum_{p, s} \bar{a}_s(p) \epsilon(p) a_s(p) - \frac{g}{\beta V} \sum_{p; i,\alpha=1,2,3} \bar{J}_{i\alpha}(p) J_{i\alpha}(p),\label{Slow}
\end{equation}
where
\begin{eqnarray}
&&p = (\omega, k), \quad \hat{k} = \frac{k}{|k|}, \\&&
\epsilon(p) = i\omega - v_F (|k|-k_F)\nonumber\\
&&J_{i\alpha}(p) = \frac{1}{2}\sum_{p_1+p_2 = p}(\hat{k}^i_1 - \hat{k}^i_2)  a_A(p_2) [\sigma_{\alpha} ]_{B}^{C} a_{C}(p_1)\epsilon^{AB}\nonumber
\end{eqnarray}

Here $V$ is the $3D$ volume, while $\beta = 1/T$ is the imaginary time extent of the model (i.e. the inverse temperature). Both $\beta$ and $V$ should be set to infinity at the end of the calculations.   $a_{\pm}(p)$ is the fermion variable in momentum space, $v_F$ is Fermi velocity, $k_F$ is Fermi momentum, $g$ is the effective coupling constant. Since the spin-orbit coupling in liquid $^3$He (the dipole-dipole interaction) is
relatively small, the spin and orbital rotation groups,
$SO_3^{ S}$ and $SO_3^{L}$, can be considered independently, and one has
\begin{equation}
G=U(1)\times SO_3^{L}\times SO_3^{ S} \,.
 \label{Bsymmetrygroup}
\end{equation}
Let us call this $G$ the high-energy symmetry. Eq. (\ref{Slow}) is invariant under the action of this group.

Next \cite{He3} we proceed with the bosonization.  The unity is substituted into the functional integral that is represented as
\begin{equation}
1 \sim \int D \bar{A} D A\,  {\rm exp}\Big(\frac{1}{g}\sum_{p, i, \alpha} \bar{A}_{i, \alpha}(p) A_{i, \alpha}(p)\Big),
\end{equation}
 where $A_{i, \alpha}, (i,\alpha = 1,2,3)$ are bosonic variables.  These variables may be considered as the field of the Cooper pairs, which serves as the analog of the Higgs field in relativistic theories.  Shift of the integrand in $D \bar{A} D A$ removes the $4$ - fermion term. Therefore, the fermionic integral can be calculated. As a result we arrive at the "hydrodynamic" action for the Higgs field $A$:
\begin{equation}
S_{eff} = \frac{1}{g} \sum_{p, i, \alpha} \bar{A}_{i, \alpha}(p) A_{i, \alpha}(p) + \frac{1}{2}{\rm log} \, {\rm Det} M(\bar{A},A), \label{hydrodynamic}
\end{equation}
where
\begin{widetext}
\begin{equation}
M(\bar{A},A) = \left(\begin{array}{cc}(i\omega - v_F (|k|-k_F))\delta_{p_1 p_2} & \frac{1}{(\beta V)^{1/2}}[(\hat{k}^i_1 - \hat{k}^i_2)A_{i\alpha}(p_1+p_2)]\sigma_{\alpha}\\-\frac{1}{(\beta V)^{1/2}}[(\hat{k}^i_1 - \hat{k}^i_2)\bar{A}_{i\alpha}(p_1+p_2)]\sigma_{\alpha}& - (i\omega - v_F (|k|-k_F))\delta_{p_1 p_2}\end{array}\right) \label{matrix}
\end{equation}
\end{widetext}

The  relevant symmetry group $G$ of the physical laws,
which is  broken in superfluid  phases of $^3$He, contains
the group $U(1)$, which is responsible for conservation of the particle number,
 and the group of rotations  $SO_3^{J}$.
This symmetry is spontaneously  broken in superfluid  phases of $^3$He.
The order parameter -- the
high-energy Higgs field -- belongs to the representation $S=1$ and $L=1$ of the
$SO_3^{S}$ and
$SO_3^{L}$ groups and is represented by $3\times 3$ complex matrix
$A_{i\alpha}$ with 18 real components.

\subsection{Vacuum of $^3$He-B }
\label{B-phase}

\label{BbrokenRelativeSymmetry}

In  superfluid $^3$He-B, the $U(1)$ symmetry and the relative spin-orbit
symmetry are broken, and the vacuum states are determined by the phase
$\Phi$ and by the rotation (orthogonal) matrix $R_{i\alpha}$:
\begin{equation}
A^{(0)}_{i\alpha} \sim
\Delta ~e^{i\Phi}~R_{i\alpha}\,.
 \label{Aalphai}
\end{equation}
Here $\Delta$ is the gap in the spectrum of fermionic quasiparticles.
The symmetry  $H$ of the vacuum state is the diagonal $SO_3$ subgroup of $G$:
the vacuum state is invariant under combined rotations.
Space $\mathcal{R}$ of the degenerate vacuum states
in  $^3$He-B includes the circumference $U(1)$ of
the phase $\Phi$ and the $SO_3$ space
of the relative rotations:
\begin{equation}
{\mathcal{R}}=G/H=U(1)\times SO_3\,.
 \label{Relative}
\end{equation}

The number of the Nambu-Goldstone modes in this symmetry breaking scenario
is $7-3=4$, while the other 14 collective modes of the order parameter
$A_{\alpha i}$ are Higgs bosons.
These 18 bosons satisfy the Nambu sum rule, which relates the masses of
bosonic and fermionic excitations \cite{Nambu1985}. The possible extension of
this rule to the Standard Model Higgs bosons is discussed in
Ref. \cite{VolovikZubkovHiggs,VolovikZubkov2014}.


In the B - phase of $^3$He the condensate is formed in the state with
$J=0$, where ${\bf J}= {\bf L}+{\bf S}$ is the total angular momentum of Cooper pair \cite{VollhardtWolfle1990}. \revision{In the absence of spin - orbit interactions matrix $R_{i\alpha}$ may be absorbed within Eqs. (\ref{hydrodynamic}), (\ref{matrix}) by the rotation of vector $k^i$. At the same time the phase $\Phi$ may be absorbed by the transformation $M(\bar{A},A) \rightarrow {\rm diag}(e^{2 i \Phi}, e^{-2i\Phi})\, M(\bar{A},A)\, {\rm diag}(e^{-2 i \Phi}, e^{2i\Phi})$ that does  not change the value of the determinant in Eq. (\ref{hydrodynamic}). As a result the vacuum is invariant under the combined spin and orbit rotations.}
So, we consider the state
\begin{equation}
A^{(0)}_{i \alpha}(p) = (\beta V)^{1/2} \frac{\Delta}{2} \, \delta_{p0}\delta_{i\alpha} \,
 \label{SymmetricVacuum}
\end{equation}
 as the symmetric low-energy vacuum.  Parameter $\Delta$ satisfies gap equation
\begin{equation}
 0=\frac{3}{g}- \frac{4}{\beta V} \sum_p (\omega^2 + v_F^2(|k|-k_F)^2+\Delta^2 )^{-1}\label{gap}
\end{equation}

$\Delta $ is the constituent mass of the fermion excitation.
We denote the fluctuations around the condensate by $\delta A_{i \alpha}=A_{i \alpha}- A^{(0)}_{i \alpha}$. Tensor $\delta A_{i\alpha}$ realizes the reducible representation of the $SO_J(3)$ symmetry group of the vacuum (acting on both spin and orbital indices).  The mentioned modes are classified by the total angular momentum quantum number $J = 0,1,2$.

\subsection{Collective modes in $^3$He-B}

According to \cite{He3gauss,He3B} the quadratic part of the effective action for the fluctuations around the condensate has the form:
\begin{equation}
S^{(1)}_{eff} =\frac{1}{g} (u,v) [1 - g \Pi]   \left(\begin{array}{c}{u}\\v \end{array}\right),
\end{equation}
where $\delta A_{i\alpha}(p) = u_{p i \alpha} + i v_{p i \alpha }$, while $\Pi$ is polarization operator.
At each value of $J=0,1,2$ the  modes $u$ and $v$ are orthogonal to each other and correspond to different values of the bosonic energy gaps. The spectrum of the quasiparticles is obtained at the zeros of expressions for $\frac{\delta^2}{\delta u_{i\alpha}\delta u_{j\beta}} S^{(1)}_{eff}$ and $\frac{\delta^2}{\delta v_{i\alpha}\delta v_{j\beta}} S^{(1)}_{eff}$. The energy gaps appear \cite{He3B} as  the solutions of equation ${\rm Det}\, \Bigl(g\Pi(i E) - 1\Bigr) = 0$:
\begin{equation}
E_{u,v}^{(J)} = \sqrt{ 2 \Delta^2(1\pm  \eta^{(J)})}  \,,
\end{equation}
This  proves the Nambu sum rule for $^3$He-B \cite{Nambu1985,VolovikZubkovHiggs,VolovikZubkov2014}:
\begin{equation}
[E_{u}^{(J)}]^2 + [E_{v}^{(J)}]^2  = 4 \Delta^2\label{NSRThe}
\end{equation}
Explicit calculation gives  $\eta^{J=0} = \eta^{J=1} = 1$, and $\eta^{J=2} = \frac{1}{5}$.  The 18 collective modes (9 real and 9 imaginary deviations $\delta A_{\alpha
i}$ of the high-energy
order parameter from the vacuum state Eq. (\ref{SymmetricVacuum})),
decompose under the $SO_3^{J} $ group as
\begin{equation}
J=0^-,  J=1^+, J=0^+, J=1^-, J=2^{\pm}\,,
 \label{decomposition}
\end{equation}
Here $+$ and $-$ correspond to real and imaginary perturbations $\delta
A_{\alpha i}$.
The bosons in the first two representations are NG bosons in the
absence of spin-orbit coupling:
the first one is the sound mode, which appears due to broken $U(1)$
symmetry;
and the second set represents three spin wave modes.

The other sets represent $1+3+5+5=14$ heavy Higgs amplitude modes with energies of
order of fermionic gap $\Delta$. These are: the so-called pair breaking
mode with $J=0^+$ and mass $2\Delta$; three pair breaking
modes with $J=1^-$  and mass $2\Delta$;  five the so-called real squashing
modes with $J=2^+$ and mass $\sqrt{12/5}\Delta$; and  five imaginary
squashing modes with $J=2^-$ and mass $\sqrt{8/5}\Delta$.

\subsection{Taking into account the spin-orbit interactions}

\revision{The spin-orbit interaction reduces the degeneracy of the vacuum space and
transforms one of the NG modes to the massive Higgs boson.
Under the spin-orbit interaction the high-energy symmetry group $G$ is
reduced to the low-energy symmetry group
\begin{equation}
G_{\rm so}=U(1)\times SO_3^{J} \,,
 \label{BsymmetrySO}
\end{equation}
where $SO_3^{J} $ is the group of combined rotations in spin and orbital
spaces. The spin - orbit interaction gives the following contribution to the effective low energy action \cite{VollhardtWolfle1990}:
\begin{eqnarray}
S_{SO}[A] & = & \frac{3}{5}g_D \sum_{p}\bar{A}_{i, \alpha}(p) A_{j, \beta}(p)\Big(\delta_{i\alpha}\delta_{j\beta}+\delta_{j\alpha}\delta_{i\beta}\nonumber\\&&-\frac{2}{3}\delta_{ij}\delta_{\alpha\beta} \Big),\label{SSO}
\end{eqnarray}
where $g_D$ is the new coupling constant. Matrix $R_{i,\alpha}$ still can be absorbed by the rotation of $k^i$ in Eq. (\ref{matrix}). However, the complete effective action depends on it due to the contribution of Eq. (\ref{SSO}).   As a result instead of Eq. (\ref{SymmetricVacuum}) we keep
\begin{equation}
A^{(0)}_{i \alpha}(p) = (\beta V)^{1/2} \frac{\Delta}{2} \, \delta_{p0}R_{i\alpha}, \,
 \label{ASymmetricVacuum}
\end{equation}
where
orthogonal matrix $R_{i\alpha}$ may be represented in terms
of the angle $\theta$ and the axis $\hat{\bf n}$ of rotation:
\begin{equation}
R_{i \alpha}(\hat{\bf n}, \theta)=\hat n_{\alpha}\hat n_i+(\delta_{\alpha
i}-\hat n_{\alpha}\hat n_i)\cos \theta- e_{\alpha ik}\hat n_k\sin \theta\,.
 \label{Ralphai}
\end{equation}
Here $\theta$ changes from $0$ to $\pi$; the points $(\hat{\bf n},  \theta=\pi)$
and $(-\hat{\bf n},  \theta=\pi)$ are equivalent.
Being substituted to Eq. (\ref{SSO}) the condensate of the form of Eq. (\ref{ASymmetricVacuum}) gives
\begin{equation}
S_{SO}[A^{(0)}]= g_D  \Delta^2 \Big(\frac{6}{5}({\rm cos}\,\theta + 1/4)^2 - \frac{3}{8}\Big)\, \beta V,\label{SSO1}
\end{equation}
Minimum of this expression is achieved, when $\theta = \theta_0 \approx 104^{\circ}$ (the so - called Leggett angle).}

\revision{In principle, Eq. (\ref{SSO}) affects the gap equation. The functional form of the condensate is given by Eq. (\ref{gap}). However, the constant $g$ entering this equation receives small $\Delta$ - dependent contribution. We neglect this sontribution in the following. The most valuable effect of the spin - orbit interaction is the appearance of the explicit mass term for the collective mode given by the fluctuations of $\theta$ around its vacuum value given by the Leggett angle $\theta_0$.}

It is worth mentioning that the interaction term of the form of Eq. (\ref{SSO}) is equivalent to a certain modification of the original four - fermion interaction of Eq. (\ref{Slow}). The modified four - fermion interaction is obtained as a result of Gaussian integration over $A_{i\alpha}$ in the functional integral.

\subsection{Higgs $\#$15 from spin-orbit interaction}

\revision{Let us consider the collective mode $\delta \theta = \theta - \theta_0$. It originates from the modes with $J=1^+$ and forms the low-energy  Higgs field --
the light Higgs.
The $J=1^+$ collective mode is the 3-vector field, whose components can be obtained from the
orthogonal matrix $R_{\alpha i}$, when it is represented in terms
of the angle $\theta$ and the axis $\hat{\bf n}$ of rotation. The directions of unit vector $\hat{\bf n}$ correspond to the two massless Goldstone modes. The field $\delta \theta$ represents  gapped collective mode. }

\revision{The mass term for this collective mode is given completely by the form of Eq. (\ref{SSO}) because the dynamical contribution coming from the integration over fermions vanishes. However, the kinetic term comes from the integration over fermions.
We represent the effect of the fluctuation $\delta \theta$ on the condensate function as follows
\begin{equation}
A_{i,\alpha}[\delta\theta] = R_{i \alpha}(\hat{\bf n}, \theta) = R_{i \alpha}(\hat{\bf n}, \theta_0)R_{i \alpha}(\hat{\bf n}, \delta \theta)
\end{equation}
Within the functional determinant we absorb $R_{i \alpha}(\hat{\bf n}, \theta_0)$ by the rotation of $k^i$. The remaining part gives actual form of $\delta A_{i,\alpha}$:
\begin{equation}
\delta A_{i,\alpha} = - e_{\alpha ik}\hat n_k\, \delta \theta\,(\beta V)^{1/2} \frac{\Delta}{2} \,
\end{equation}
The kinetic term for $\delta \theta$ has the form $S_{\rm kin}[\delta \theta] = \sum_{\omega,k}\Pi_{\theta}(\omega,k)[\delta \theta(\omega,k)]^2$, where
\begin{eqnarray}
\Pi_{\theta}(\omega,0)& = & -\frac{1}{4} \sum_{\epsilon,k}\,{\rm Sp}\, G(\epsilon+\omega,k) \, O(\hat{n})\, G(\epsilon,k)\, O(\hat{n}) \nonumber\\ &\approx & Z^2_{\theta} \omega^2\label{Zo}
\end{eqnarray}
with
\begin{equation}
G^{-1}(\epsilon,k)  = \left(\begin{array}{cc} (i\epsilon-v_F(|k|-k_F)) & \Delta (\hat{k}\sigma)\\
-\Delta (\hat{k}\sigma)&(-i\epsilon+v_F(|k|-k_F))\end{array} \right)
\end{equation}
and
\begin{eqnarray}
&& O(\hat{n})  = \left(\begin{array}{cc} 0  & \hat{k}^i e_{i\alpha k}\sigma^{\alpha} \hat{n}^k\\
-\hat{k}^i e_{i\alpha k}\sigma^{\alpha} \hat{n}^k& 0 \end{array} \right)
\end{eqnarray}
Constant $Z_{\theta}$ enters the expression for the effective action of $\theta(\omega,0)$:
\begin{equation}
S_{\theta} \approx \sum_{\omega}\Big(Z^2_{\theta}\omega^2 + \frac{9}{4} g_D  \Delta^2 \Big)[\delta \theta(\omega,0)]^2
\end{equation}
This gives the following expression for the energy gap of the LH mode:
\begin{equation}
E_{\theta}= \Omega_B = \frac{3}{2 Z_{\theta}} \sqrt{g_D} \Delta
\end{equation}
Here $\Omega_B$ is the Leggett frequency (the frequency of the longitudinal NMR) in $^3$He-B
\cite{VollhardtWolfle1990}.}

\revision{In the language of quantum field theory $Z^2_{\theta}$ is the wave function renormalization constant for the field $\theta$. It depends logarithmically on the width of the region of momenta around the Fermi surface. This is the region over which we should integrate in Eq. (\ref{Zo}). Using manipulations with the derivatives of the partition function we are able to relate $Z_{\theta}$ with spin susceptibility $\chi_B = \frac{d}{dB}\langle \sigma\rangle$, where $\langle \sigma \rangle$ is the spin density in the presence of magnetic field $B$:
\begin{equation}
\chi_B = \gamma^2 Z_{\theta}^2
\end{equation}
Here ${\gamma}$ is the gyromagnetic
ratio  for the $^3$He atom.}
This allows to rewrite the $\theta$ dependent part of Eq. (\ref{SSO1}) for the spin - orbit interaction as
\begin{equation}
S_{SO}[\theta] =\frac{32}{15}\frac{\chi_B}{\gamma^2}\Omega_B^2(|{\bf
n}|^2 -n_0^2)^2\,\beta V,
 \label{FD}
\end{equation}
where $n_0=\sqrt{5/8}$, which corresponds to  the Leggett angle
$\cos\theta_0 =-{1\over 4}$
measured in NMR experiments.
Here we represent the field of the $J = 1^+$ collective modes (see Eqs.(2.2) and (2.3) in \cite{VolovikMineev1977}) as
\begin{equation}
{\bf n}=\hat {\bf n}\sin \frac{\theta}{2}\,.
 \label{vector}
\end{equation}

The spin-orbit interaction  fixes  the magnitude of the light Higgs field, $|{\bf
n}|=n_0$, in the equilibrium, but
leaves the degeneracy corresponding to the other two components of the $J= 1^+$ collective mode given by the direction of $\hat{\bf n}$. This
corresponds to the symmetry breaking scheme $SO_3^{J}\rightarrow SO_3^{J}/SO_2^{J}$,
where $SO_2^{J}$ is the symmetry group of rotations around axis
$\hat{\bf n}$. Thus the Higgs mechanism gives rise to two NG modes and one LH, i.e.
the spin-orbit interaction (\ref{FD}) transforms one of the NG modes to the LH mode.

The mass of the LHB is determined by the parameters in Eq.  (\ref{FD}).
The Leggett
frequency $\Omega_B$ determines the mass of the amplitude Higgs mode -- the
$\theta$-boson with the dispersion low
\begin{equation}
E^2=\Omega_B^2+ c^2k^2~~
 \label{LHiggsB}
\end{equation}
Here $c$ is the relevant speed of spin waves, which in general depends on the
direction of propagation  \cite{VollhardtWolfle1990}.
In  $^3$He-B,  $\Omega_B \sim 10^{-3}\Delta$, i.e. the
light Higgs acquires the mass, which is much lower than the energy scale
$\Delta$,
at which the symmetry breaking occurs and which characterizes the energies
of the
heavy Higgs bosons.
Note that in  $^3$He-B,  the low-energy physics has all the signatures of
the Higgs scenario. The low-energy vector Higgs field  ${\bf n}$ has both
the massive amplitude mode and two massless NG bosons.

In applied magnetic field the time reversal symmetry is violated, and  two massless NG modes
transform to the mode with the Larmor gap (magnon) and NG mode with quadratic dispersion.
The parametric decay of magnons to the pairs of the LH bosons has been recently observed in NMR experiments with Bose-Einstein condensates of magnons \cite{Zavjalov2014}.

The given scenario in $^3$He-B does not say anything on the NG mode, which
comes from the breaking of $U(1)$ symmetry. The latter is determined by the high-energy
physics and is not influenced by spin-orbit coupling.
When the spin-orbit coupling is taken into account, the symmetry breaking scheme
gives
\begin{equation}
{\mathcal{R}}_{\rm so}=G_{\rm so}/H_{\rm so}=U(1)\times
SO_3^{J}/SO_2^{J}=U(1)\times S^2\,.
 \label{SymmetryBreakingSO}
\end{equation}
This results in the $2+1$ NG bosons instead
of $3+1$  NG bosons in the absence of spin-orbit coupling.

The $U(1)$ degree of freedom does not appear if instead of
superfluid  $^3$He-B one
considers a non-superfluid  antiferromagnetic liquid crystal.
Here the  transition  occurs without breaking of $U(1)$ symmetry,
and $U(1)$ drops out of  Eqs. (\ref{BsymmetrySO}) and (\ref{Relative}).
Such transition is fully determined by the real-valued order parameter matrix
$A_{\alpha i}$.
If the relative spin-orbit symmetry is broken in the same manner
as in $^3$He-B, one obtains
in the absence of spin-orbit coupling $1+5$ heavy Higgs bosons with $J=0$
and $J=2$; and 3 NG bosons with $J=1$. The spin-orbit coupling then
transforms one of the NG bosons to the light Higgs.

\subsection{Polar phase of superfluid  $^3$He}
\label{PolarPhase}

Polar phase of superfluid  $^3$He has been recently observed in strongly
anisotropic alumina aerogel \cite{Dmitriev2012,Mineev2014}. New phases
of superfluid  $^3$He with strong polar distorsion have been also reported
in anisotropic aerogel  \cite{Dmitriev2014}. Here we neglect the
anisotropy of aerogel. Inclusion of this anisotropy is straightforward, and
does not influence the mechanism of the light Higgs mass generation.

\subsubsection{Neglected spin-orbit interaction}
\label{noSpinOrbit}

 In the  polar phase, the $U(1)$ symmetry is broken, and each of the two $SO_3$ groups is
broken to its $SO_2$ subgroup:
$H=SO_2^{S}\times SO_2^{L}$.
The order parameter matrix $A_{\alpha i}$ in the polar phase vacuum has the
form:
\begin{equation}
A_{\alpha i}=
\Delta ~e^{i\Phi}~\hat d_{\alpha} \hat m_{i}\,,
 \label{Apolar}
\end{equation}
where $\hat {\bf d}$ and $\hat {\bf m}$ are unit vectors.
Space $\mathcal{R}$ of the degenerate states
in the  polar phase includes the circumference $U(1)$ of
the phase $\Phi$ and the two $S^2$ spheres:
\begin{equation}
{\mathcal{R}}=G/H=U(1)\times S^2 \times S^2\,.
 \label{Rpolar}
\end{equation}
The high-energy polar phase has $1+2+2=5$ NG modes and $18-5=13$ heavy Higgs
modes with mass (gap) of order $\Delta$.
The anisotropy of aerogel fixes the orbital vector $\hat {\bf m}$ and thus
removes 2 NG modes.

\subsubsection{Higgs $\#$14 from spin-orbit interaction}

When the spin-orbit interaction is taken into account, the symmetry breaking
scheme becomes
\begin{equation}
G_{\rm so}=U(1)\times SO_3^{J} ~~,~~ H_{\rm so}=1~~,~~{\mathcal{R}}_{\rm
so}=G_{\rm so}\,.
 \label{SymmetryBreakingSOPolar}
\end{equation}
The spin-orbit interaction reduces the degeneracy of the vacuum space,
${\mathcal{R}}_{\rm so}<{\mathcal{R}}$, leaving only $1+3=4$ NG modes
(two of which are removed by strong orbital anisotropy of aerogel).
As a result, the spin-orbit coupling transforms one of the NG modes to the massive Higgs
boson -- the light Higgs.

Let us start with vacuum state
with $\hat {\bf d}=\hat {\bf m}=\hat {\bf z}$. This vacuum state has quantum
numbers
$S_z=L_z=0$, and thus $J_z=0$, which corresponds to symmetry $SO_2^{J}$ of
the vacuum state.
This symmetry is broken by light Higgs. The LH field can be introduced for example
as the real vector field ${\bf n}\perp \hat {\bf z}$, which describes the
deviation  $\hat {\bf d}-\hat {\bf m}$:
\begin{equation}
\hat {\bf m}=\hat {\bf z} \sqrt{1- |{\bf n}|^2} + {\bf n}~~,~~\hat {\bf
d}=\hat {\bf z} \sqrt{1- |{\bf n}|^2} - {\bf n}\,.
 \label{Psi-polar}
\end{equation}
In terms of the vector ${\bf n}$ the spin-orbit interaction in the polar phase is
\begin{equation}
F_{\rm so}=2\frac{\chi}{\gamma^2}\Omega_{\rm pol}^2(|{\bf
n}|^2 -n_0^2)^2\,,
 \label{FDpolar}
\end{equation}
where $\Omega_{\rm pol}$ is the Leggett frequency for the polar phase, and $n_0=\sqrt{1/2}$.
The spin-orbit interaction  fixes  the magnitude of the little Higgs field $|{\bf
n}|$ in the equilibrium, but
leaves the degeneracy with respect to its orientation
in the plane perpendicular to $z$-axis.
This leads to one NG boson -- the spin wave mode with spectrum $E=cp$, and the
light Higgs mode:
\begin{equation}
E^2=\Omega_{\rm pol}^2+ c^2k^2\,,
 \label{LHiggsPolar}
\end{equation}
with mass (gap) $\Omega_{\rm pol}\ll \Delta$.

\section{A model with the pseudo - Goldstone boson composed of the top quark}
\label{seesaw}

\subsection{Dynamical symmetry breaking and dynamical masses of quarks}

\subsubsection{Lagrangian}

Let us consider the model inspired by the top seesaw model suggested by Cheng, Dobrescu and Gu in
\cite{dobrescu}. This model contains (in addition to the SM fermions) the
 fermion $\chi$. The action
contains the four - fermion interaction terms, that being written
through the auxiliary $3$ - component field $\Phi$ have the form:
\begin{eqnarray}
L_I &=& - M_0^2 \Big(\frac{1}{\xi^2_t} \Phi_t^+ \Phi_t +
\frac{1}{\xi^2_{\chi}} \Phi_{\chi}^+ \Phi_{\chi}\nonumber\\&& +
\frac{1}{\xi^2_{t\chi}} [\Phi_t^+
\Phi_{\chi}+\Phi_{\chi}^+\Phi_t]\Big) \nonumber\\&&- \Bigl[
\left(\begin{array}{ccc} \bar{b}^{\prime}_L & \bar{t}^{\prime}_L  & \bar{\chi}^{\prime}_L \end{array} \right)
\Phi_t t^{\prime}_R + \left(\begin{array}{ccc} \bar{b}^{\prime}_L & \bar{t}^{\prime}_L  & \bar{\chi}^{\prime}_L \end{array} \right)
\Phi_{\chi} {\chi}^{\prime}_R \nonumber\\&&+
(h.c.)\Bigr],\label{LI}
\end{eqnarray}
For the convenience of the further consideration we have changed the order of $t^{\prime}$ and $b^{\prime}$ compared to \cite{dobrescu}. Also for the convenience we denote $\Phi = (0, \Phi_t, \Phi_{\chi})$ and
\begin{equation}
L_I = -   {\rm Tr}\, \Phi  \Omega \Phi^+  - \Bigl[
\bar{\psi}_L
\Phi  \psi_R + (h.c.)\Bigr],\label{LI0}
\end{equation}
where
\begin{equation}
\psi_{L} = \left(\begin{array}{c}b^{\prime}_L\\t^{\prime}_{L} \\ {\chi}^{\prime}_L \end{array}
\right), \quad \psi_R =
 \left( \begin{array}{c}b^{\prime}_R\\t^{\prime}_R\\\chi^{\prime}_R  \end{array}\right)\label{btchi}
\end{equation}
while $\Omega$ is the corresponding $3\times 3$ matrix.
Notice, that the three components of $\psi$ are equal to the fields of $b$, $t$, and $\chi$ only in the basis, in which the mass matrix is diagonal (see below). Therefore, in Eq. (\ref{btchi}) written in arbitrary basis we do not identify $b^{\prime},t^{\prime}$ and $\chi^{\prime}$ with the actual fields of $b$ - quark, top - quark and the heavy quark $\chi$.

The global symmetry of the given lagrangian is $SU(3)_L \otimes U(1)_L\otimes U(1)_{t, R}\otimes U(1)_{\chi, R}$. Here $SU(3)_L$ corresponds to the $SU(3)$ rotations of $\psi_L$, while  the $U(1)$ parts of the global symmetry of our lagrangian correspond to the transformations $\psi_L\rightarrow e^{i \alpha} \psi_L$, $\psi_{t,R} \rightarrow e^{i \beta}\psi_{t,R}$, and $\Phi_t \rightarrow e^{i (\alpha - \beta)} \Phi_t$ (and the similar transformation for $\chi$).

The quantum numbers of $\chi^\prime_L$ and $\chi^\prime_R$ including the hypercharge (and the quantum numbers of $t^\prime_R$) are equal to the quantum numbers of the right - handed top quark. This is the doublet field $\left(\begin{array}{c}b^\prime_L\\t^\prime_L \end{array}\right)$, which is transformed under the $SU(2)_L$ SM gauge field. Therefore, the gauge interactions of the SM break the $SU(3)_L$ symmetry - the effect, which we neglect here.

Using orthogonal rotation of $t_R$ and $\chi_R$ we can always bring $\Omega$ to the
diagonal form with $1/\xi_{t\chi}=0$. We denote in this representation
\begin{equation}
\Omega^{(0)} = \left(\begin{array}{ccc}0 & 0 & 0 \\0 & \omega^{(0)}_t & 0 \\ 0 & 0& \omega^{(0)}_{\chi}
\end{array} \right) =  \left(\begin{array}{ccc}0 & 0 & 0 \\0 & 1/\xi^2_t & 0 \\0 & 0 &
1/\xi^2_{\chi} \end{array} \right)M_0^2 \label{Omega0}
\end{equation}

In \cite{dobrescu} the explicit mass term in lagrangian that breaks the $SU(3)$
symmetry down to $SU(2)$ was added:
\begin{equation}
L_M = - \mu_{\chi t} \bar{\chi}_L t_R - \mu_{\chi \chi}\bar{\chi}_L\chi_R +
(h.c.)   ,\label{lm}
\end{equation}
In addition, in \cite{dobrescu} the other contributions to the lagrangian were considered that do not originate from the four - fermion interactions. A similar construction has been considered in \cite{Yamawaki}, where the original $SU(3)$ symmetry is broken both by the additional four - fermion terms and the mass term of the form of Eq. (\ref{lm}). In our model we restrict ourselves with the four - fermion interaction terms and do not consider the explicit mass term. We introduce the following modification of the four - fermion interaction that reveals an analogy with the spin - orbit interaction of $^3$He considered in the previous section (see Eq. (\ref{SSO})).

Namely, we add the following terms to the lagrangian
\begin{eqnarray}
L_G &=&  g^{(0)}_{\chi} |\Phi^3_{\chi}|^2 + g^{(0)}_{t}|\Phi^3_t|^2 + g^{(0)}_{t\chi} \Big( \bar{\Phi}^3_{\chi}\Phi^3_{t} + (h.c.) \Big) \nonumber\\ &=&  {\rm Tr}\, \Phi \, G^{(0)} \Phi^+ \Upsilon_3, \label{gtc3}
\end{eqnarray}
and
\begin{eqnarray}
L_B &=&  -b^{(0)}_{\chi} |{\rm Im}\Phi^3_{\chi}|^2 - b^{(0)}_{t}|{\rm Im}\Phi^3_t|^2 \nonumber\\&&- 2b^{(0)}_{t\chi}( {\rm Im}{\Phi}^3_{\chi})({\rm Im} \Phi^3_{t}) \nonumber\\ &=&  \frac{1}{4}{\rm Tr}\, (\Phi-\Phi^*) \, B^{(0)} (\Phi^T-\Phi^+) \Upsilon_3, \label{gtc4}
\end{eqnarray}
where
\begin{eqnarray}
G^{(0)}&=&\left(\begin{array}{ccc}0 & 0 & 0 \\0 & g^{(0)}_t & g^{(0)}_{t\chi} \\ 0 & g^{(0)}_{t\chi}& g^{(0)}_{\chi}
\end{array} \right) ,\quad B^{(0)}=\left(\begin{array}{ccc}0 & 0 & 0 \\0 & b^{(0)}_t & b^{(0)}_{t\chi} \\ 0 & b^{(0)}_{t\chi}& b^{(0)}_{\chi}
\end{array} \right) ,\nonumber\\  \Upsilon_3 & = & \left(\begin{array}{ccc}0 & 0 & 0 \\0 & 0 & 0 \\ 0 & 0& 1
\end{array} \right)\label{G3}
\end{eqnarray}
We bring $\Omega$ to the diagonal form via orthogonal rotations of $\psi_R$. Further we choose the representation in this basis. We assume that the elements of matrices $\Omega$, $B$ and $G$ are real - valued.

\subsubsection{Effective action for scalar bosons}

Let us choose the parametrization in which the massless $b$ - quark is identified with $b^{\prime}=\psi^1$. It corresponds to  the representation $\Phi = \langle \Phi \rangle + \tilde{\Phi} = V + \tilde{\Phi}$, where
\begin{eqnarray}
&& \hat{V} = \left(\begin{array}{ccc}0&0&0\\ 0&  \frac{1}{\sqrt{2}} v_t & \frac{1}{\sqrt{2}} v_{\chi} \\
0 & \frac{1}{\sqrt{2}} u_t &  \frac{1}{\sqrt{2}} u_{\chi}
\end{array}\right), \nonumber\\ && \tilde{\Phi} =
\left(\begin{array}{ccc} 0 & H_t^- &  H_{\chi}^- \\0 & \frac{1}{\sqrt{2}}( h_t +i
A_t) &  \frac{1}{\sqrt{2}}(h_{\chi} +i
A_{\chi})\\
 0 &  \frac{1}{\sqrt{2}}(\varphi_{t} +i \pi_{t}) &   \frac{1}{\sqrt{2}}(\varphi_{\chi} +i \pi_{\chi})
\end{array}\right)
\end{eqnarray}
This expression is similar to that of Eq. (2.11) in \cite{dobrescu}. Here the values of $v_{t,\chi}$ and $u_{t,\chi}$ correspond to the condensate.

Effective action for the field $\tilde{\Phi}$ has the form:
\begin{eqnarray}
S[\tilde{\Phi}] &=&  -   \int d ^4 x {\rm Tr}\, (\hat{V}+\tilde{\Phi})\Omega^{(0)} (\hat{V}+\tilde{\Phi})^+ \nonumber\\ && +   \int d ^4 x {\rm Tr}\, (\hat{V}+\tilde{\Phi})G^{(0)} (\hat{V}+\tilde{\Phi})^+\Upsilon_3  \nonumber\\&&
   +\int d ^4 x \frac{1}{4}{\rm Tr}\, (V-V^* + \Phi-\Phi^*) \,\nonumber\\&& B^{(0)} (V^T - V^+ + \Phi^T-\Phi^+) \Upsilon_3  \nonumber\\&&- i \, {\rm log}\, {\rm Det}\, \Big(i \gamma \partial - {\cal Q}\Big(\hat{V} + \tilde{\Phi}\Big) \Big)\label{SEFF}
\end{eqnarray}
Here for any matrix $O$ we define
\begin{equation}
{\cal Q} O  = \left(\begin{array}{cc}O^+ & 0\\ 0 & O
\end{array}\right)
\end{equation}
$\hat{V}+ $ plays the role of mass matrix, and we denote $\hat{m} = \hat{V} $.

\subsubsection{Gap equation}

Gap equation appears as
\begin{equation}
\frac{\delta}{\delta  \tilde{\Phi}_{ia}}S[\tilde{\Phi}] = 0, \quad i = 1,2,3,\quad a = 2,3
\end{equation}
We represent the determinant in Eq. (\ref{SEFF}) as follows
\begin{eqnarray}
&& - i \, {\rm log}\, {\rm Det}\, \Big(i \gamma \partial - {\cal Q}\Big(\hat{V}  + \hat{\mu}^{(0)} + \tilde{\Phi}\Big) \Big)\nonumber\\&&  =   {\rm const}  - i \, {\rm Sp}\, {\rm log}\, \Big(i \partial \Sigma - {\cal T}\hat{m}  \Big)\nonumber\\ && + i \, {\rm Sp}\,  \frac{1}{i \partial \Sigma - {\cal T}\hat{m}}{\cal T} \tilde{\Phi}\nonumber\\&& + \frac{i}{2} \, {\rm Sp}\,  \frac{1}{i \partial \Sigma - {\cal T}\hat{m}}{\cal T} \tilde{\Phi}\,  \frac{1}{i \partial \Sigma - {\cal T}\hat{m}}{\cal T} \tilde{\Phi}  + ...
\end{eqnarray}
Here
\begin{equation}
\Sigma = \left(\begin{array}{cc} \bar{\sigma} & 0 \\ 0 & {\sigma} \end{array} \right),
\quad {\cal T}O = \gamma^0 {\cal Q}O= \left(\begin{array}{cc}0 & O \\ O^+ & 0
\end{array}\right)
\end{equation}

This gives for the gap equation ($ i = 2,3$ and $a = 2,3$).
\begin{widetext}
\begin{equation}
\Big[\Omega^{(0)} \hat{V}^+ +  (i\,B\, {\rm Im} V-G^{(0)} \hat{V}^+) \Upsilon_3\Big]_a^i= \frac{2i}{(2\pi)^4} \int \Big[\frac{d^4 p}{p^2 - \hat{m}^+\hat{m}} \hat{m}^+\Big]^i_a = -\langle\bar{\psi}^i_L \psi_{a,R} \rangle
\end{equation}
\end{widetext}
First of all, Eq. (\ref{gtc4}) suppresses the imaginary parts of $\Phi_{i\alpha}$. Therefore, this is reasonable to look for the solutions of the gap equation with real - valued $\hat{V}$.
This allows to eliminate matrix $B$ from the consideration of gap equations:
\begin{equation}
\Omega^{(0)} \hat{m}^+ -  G^{(0)} \hat{m}^+\Upsilon_3= \frac{N_c}{8\pi^2}  \Big( \Lambda^2 - \hat{m}^+\hat{m} \, {\rm log} \frac{\Lambda^2}{\hat{m}^+\hat{m}}\Big)\hat{m}^+\label{gapeq}
\end{equation}
Let us perform orthogonal rotations of $\psi_{L,R}$ that bring $\hat{m}$
to the diagonal form:
\begin{eqnarray}
&& \psi_L \rightarrow \Theta \psi_L, \quad \psi_R \rightarrow A \psi_R,\nonumber\\ && \hat{m} \rightarrow \Theta^T \hat{m} A = {\rm diag}(0,m_t,m_{\chi})\label{rot}
\end{eqnarray}
where
\begin{eqnarray}
\Theta & = & {\rm exp}\, \Big(-i \theta \sigma^2 \Big), \quad A  = {\rm exp}\,
\Big(-i \alpha \sigma^2 \Big), \nonumber\\ \sigma_2 & = & \left(\begin{array}{ccc}1&0&0\\ 0& 0  & -i  \\
0 & i  & 0
\end{array}\right)\label{rot2}
\end{eqnarray}
As a result we come to the following form of gap equation with diagonal matrix $\hat{m}$:
\begin{eqnarray}
&& A^T \Omega^{(0)} A   - A^T\, G^{(0)} \, A \, \hat{m} \Theta^T \, \Upsilon_3 \, \Theta \,\hat{m}^{-1}  \nonumber\\ &&= \frac{N_c}{8 \pi^2}
\Big(\Lambda^2 - \hat{m}^2 \, {\rm log} \, \frac{\Lambda^2}{\hat{m}^2}\Big),\label{gap4}
\end{eqnarray}

We assume, that the $SU(3)$ breaking terms are small, that is
\begin{equation}
\frac{g^{(0)}_{t,\chi,t\chi}}{\omega^{(0)}_{t,\chi}} \ll 1
\end{equation}
This does not mean, however, that the resulting corrections to fermion and boson masses are small if we consider the system near to the criticality and disregard the next to leading $1/N_c$ corrections (see discussion in the Introduction).

We also assume $m_t\ll m_{\chi}$ and  $\theta \ll 1$.
By $g_{t,\chi}$ we denote the elements of matrix $A^T\, G \, A$ that are related to the original parameters $g^{(0)}_{t,\chi}$ as follows:
\begin{eqnarray}
 g_{t} & = &  ({\rm cos}\, \alpha\,\,g^{(0)}_{t}+{\rm sin}\, \alpha\,\,g^{(0)}_{t\chi})\,{\rm cos}\, \alpha\,\nonumber\\ &&+({\rm cos}\, \alpha\,\,g^{(0)}_{t\chi}+{\rm sin}\, \alpha\,\,g^{(0)}_{\chi})\,{\rm sin}\, \alpha\, \nonumber\\
g_{t\chi}& = & -({\rm cos}\, \alpha\,\,g^{(0)}_{t}+{\rm sin}\, \alpha\,\,g^{(0)}_{t\chi})\,{\rm sin}\, \alpha\, \nonumber\\ &&+({\rm cos}\, \alpha\,\,g^{(0)}_{t\chi}+{\rm sin}\, \alpha\,\,g^{(0)}_{\chi})\,{\rm cos}\, \alpha\, \nonumber\\ g_{\chi} &=&  -(-{\rm sin}\, \alpha\,\,g^{(0)}_{t}+{\rm cos}\, \alpha\,\,g^{(0)}_{t\chi})\,{\rm sin}\, \alpha\,\nonumber\\ &&+(-{\rm sin}\, \alpha\,\,g^{(0)}_{t\chi}+{\rm cos}\, \alpha\,\,g^{(0)}_{\chi})\,{\rm cos}\, \alpha\,\label{G3A}
\end{eqnarray}
Direct calculation gives the following relation between the angle $\theta$, the ratio $m_t/m_{\chi}$, and the values of $g_{t,\chi}$:
\begin{eqnarray}
0 & = & (g_{t}\,m_{t}\,{\rm sin}\,\theta\,+g_{t\chi}\,m_{\chi}\,{\rm cos}\,\theta\,)\,{\rm cos}\,\theta\,/m_{\chi}\nonumber\\&&-(g_{t\chi}\,m_{t}\,{\rm sin}\,\theta\,+g_{\chi}\,m_{\chi}\,{\rm cos}\,\theta\,)\,{\rm sin}\,\theta\,/m_{t}\label{theta}
\end{eqnarray}
Therefore,
\begin{equation}
\theta \approx \frac{g_{t\chi}}{g_{\chi}-\frac{m_t^2}{m^2_{\chi}}g_t}\frac{m_t}{m_{\chi}} + O(m_t^3)
\end{equation}
For the angle $\alpha$ we have
\begin{eqnarray}
 &&\omega_{t\chi} \equiv  \frac{1}{2}(\omega^{(0)}_{\chi}-\omega^{(0)}_t)\, {\rm sin}\, 2 \alpha \nonumber\\&& = \Big( g_t\, \frac{m_t}{m_{\chi}} \, {\rm sin}\,\theta +g_{t\chi} \, {\rm cos}\, \theta \Big)\, {\rm cos}\, \theta  \approx g_{t\chi} \label{alpha}
\end{eqnarray}
This leads to
\begin{equation}
\alpha \approx \frac{1}{2}{\rm arctg}\, \frac{2g^{(0)}_{t\chi}}{\omega^{(0)}_{\chi}-\omega^{(0)}_t-g^{(0)}_{\chi}+g^{(0)}_t} + O(m^2_t)
\end{equation}

We are left with the following
equations:
\begin{eqnarray}
&&{\omega}_{t} - f_t   =  \frac{N_c }{8 \pi^2}\,\Big(\Lambda^2 - m_t^2
\, {\rm
log} \frac{\Lambda^2}{m_t^2}\Big); \nonumber\\ && {\omega}_{\chi}-f_{\chi}  =
\frac{N_c}{8 \pi^2}\,\Big(\Lambda^2 - m_{\chi}^2 \, {\rm log}
\frac{\Lambda^2}{m_{\chi}^2}\Big),\label{gapeqs}
\end{eqnarray}
where $\Lambda$ is the ultraviolet cutoff (of the order of the scale of the new hidden interaction), while
\begin{eqnarray}
{\omega}_{t,\chi} &=& {\rm cos}^2 \alpha \, \omega^{(0)}_{t,\chi} + {\rm sin}^2\alpha\, \omega^{(0)}_{\chi,t}\label{omegas}
\end{eqnarray}
and
\begin{eqnarray}
f_t & = &{\rm sin} \, \theta\, \Big(g_t {\rm sin}\, \theta+g_{t\chi} \frac{m_{\chi}}{m_t}\, {\rm cos}\, \theta\Big) \approx \frac{g^2_{t\chi}}{g_{\chi}} + O(m_t^2),\nonumber\\ f_{\chi} &=& {\rm cos} \, \theta\, \Big(g_{t\chi}\frac{m_t}{m_{\chi}} {\rm sin}\, \theta+g_{\chi} \, {\rm cos}\, \theta\Big)\approx g_{\chi} +O(m_t^2)\nonumber
\end{eqnarray}
Gap equation provides that $\omega^{(0)}_{t,\chi} \sim \frac{N_c}{8 \pi^2}\, \Lambda^2$ while $\omega^{(0)}_{\chi}-\omega^{(0)}_t \sim m^2_{\chi}$. Therefore, in general case $\alpha$ is not small.

For the calculation of the scalar boson spectrum we will need the exact expressions for $f_t , f_{\chi}$ through $\theta$ and the exact expression that relates $m_t^2/m_{\chi}^2$ and $\theta$. In the following we shall use in our expressions the values of $g_{t, \chi, t\chi}$ but we should remember that they differ from the original parameters $g^{(0)}_{t,\chi,t\chi}$. In principle, Eqs. (\ref{G3A}) and (\ref{theta}) allow to determine precisely $\theta$ and $\alpha$ as functions of $g^{(0)}_{t,\chi,t\chi}$ and then $g_{t,\chi,t\chi}$ as functions of $g^{(0)}_{t,\chi,t\chi}$. However, the corresponding expressions are so complicated that we do not represent them here.

\subsection{Effective action for scalar bosons}

\subsubsection{Polarization operator}

Let us consider the system in the
parametrization, in which the fermion mass matrix is diagonal. Those fermion fields that are the mass eigentstates are expressed linearly through the original fields $t^\prime_L$, $\chi^\prime_L$, $t^\prime_R$, $\chi_R^\prime$. This is the doublet field $\left(\begin{array}{c}b^\prime_L\\t^\prime_L \end{array}\right)$, which is transformed under the $SU(2)_L$ SM gauge field. At the same time $\chi^\prime_L$ has the quantum numbers of $t_R$. Thus, the mass eigenstates do not have definite charges with respect to the SM gauge fields.   Below we neglect the influence of the gauge fields on dynamics of the scalar bosons. We shall consider the terms in effective action with the interaction between the gauge fields of the Standard Model and the composite scalar bosons in Section \ref{SectPhenomenology}.

In this basis $\Omega$
has the form
\begin{eqnarray}
\Omega &=& A^T \, {\rm diag} (\omega^{(0)}_t, \omega^{(0)}_{\chi}) \, A =
\left(\begin{array}{cc}{\omega}_t & \omega_{t\chi}\\ \omega_{t\chi} & {\omega}_{\chi} \end{array}\right),  \nonumber\\&& \omega^2_{t\chi} = f_t f_{\chi}  \label{OMEGAf}
\end{eqnarray}
In the same way we substitute $G = A^T G^{(0)} A$, $B=A^T B^{(0)} A$ and $\Upsilon = \Theta^T \Upsilon_3 \Theta$ instead of $G^{(0)}$, $B^{(0)}$, and $\Upsilon_3$.

Taking into account that $\frac{\delta}{\delta \tilde{\Phi} }S[\tilde{\Phi}]=0$ we come to
\begin{eqnarray}
S[\tilde{\Phi}] & = &   -\int d ^4 x {\rm Tr}\, \tilde{\Phi}\Omega \tilde{\Phi}^+ + \int d ^4 x {\rm Tr}\, \tilde{\Phi} G \tilde{\Phi}^+\Upsilon \nonumber\\&& +   \int d ^4 x \frac{1}{4}{\rm Tr}\, (\Phi-\Phi^*) \, B (\Phi^T-\Phi^+) \Upsilon \nonumber\\&& - i \, {\rm Sp}\, {\rm log}\, \Big(i \gamma \partial - \hat{m}  \Big) \nonumber\\&& + \frac{i}{2} \, {\rm Sp}\,  \frac{1}{i \gamma \partial - \hat{m}} {\cal Q}\tilde{\Phi}\,  \frac{1}{i \gamma \partial- \hat{m}}{\cal Q} \tilde{\Phi}  + ...\label{SEFF}
\end{eqnarray}

 Let us denote $\Phi(p) = \int d^4 x \Phi(x) e^{i p x}$, and  $\tilde{\Phi}_{ia}(p) = \tilde{\Phi}^{\prime}_{ia} (p)+ i \tilde{\Phi}^{\prime\prime}_{ia}(p)$. The CP - even scalar states are given by the real parts of the components of $\Phi(p)$ while imaginary parts correspond to the CP - odd states. Then we have $S = {\rm const} + S^{\prime} + S^{\prime\prime}$ with
\begin{widetext}
\begin{eqnarray}
S^{\prime}[\tilde{\Phi}] & \approx &    -\sum_{abi}\int \frac{d^4 p}{(2\pi)^4} \, \tilde{\Phi}^{\prime}_{ia}(p)\Omega_{ab} \tilde{\Phi}^{\prime}_{ib}(p) +\sum_{abij}\int \frac{d^4 p}{(2\pi)^4} \, \tilde{\Phi}^{\prime}_{ia}(p)G_{ab} \tilde{\Phi}^{\prime}_{jb}(p)\Upsilon^{ij}  \\ &&  + \int \frac{d^4 p}{(2\pi)^4} \sum_{ai}\frac{2iN_c}{(2\pi)^4} \, \int  \frac{d^4 k }{(k^2 - m_i^2)((k+p)^2 - m_a^2)}\Big(k(p+k)[\Phi^{\prime}_{ia}(p)]^2 + m_im_a \Phi^{\prime}_{ai}(p)\Phi^{\prime}_{ia}(p)  \Big)  \nonumber\\
S^{\prime\prime}[\tilde{\Phi}]  & \approx &    -\sum_{abi}\int \frac{d^4 p}{(2\pi)^4} \, \tilde{\Phi}^{\prime\prime}_{ia}(p)\Omega_{ab} \tilde{\Phi}^{\prime\prime}_{ib}(p) +\sum_{abij}\int \frac{d^4 p}{(2\pi)^4} \, \tilde{\Phi}^{\prime\prime}_{ia}(p)G_{ab} \tilde{\Phi}^{\prime\prime}_{jb}(p)\Upsilon^{ij}  \\ &&  -\sum_{abij}\int \frac{d^4 p}{(2\pi)^4} \, \tilde{\Phi}^{\prime\prime}_{ia}(p)B_{ab} \tilde{\Phi}^{\prime\prime}_{jb}(p)\Upsilon^{ij}\nonumber\\&&+ \sum_{ai}\int \frac{d^4 p}{(2\pi)^4} \frac{2iN_c}{(2\pi)^4} \, \int  \frac{d^4 k }{(k^2 - m_i^2)((k+p)^2 - m_a^2)}\Big(k(p+k)[\Phi^{\prime\prime}_{ia}(p)]^2 - m_im_a \Phi^{\prime\prime}_{ai}(p)\Phi^{\prime\prime}_{ia}(p)  \Big)  \nonumber
\end{eqnarray}
\end{widetext}

Masses of scalar bosons appear as the zeros of operators
\begin{eqnarray}
{\cal P}^{\prime}_{(ia)(jb)}(p)& = & -(2\pi)^4\frac{\delta^2}{\delta \tilde{\Phi}^{\prime}_{ia}(p) \delta \tilde{\Phi}^{\prime}_{jb}(p) }S[\tilde{\Phi}], \nonumber\\  {\cal P}^{\prime\prime}_{(ia)(jb)}(p) &=& -(2\pi)^4 \frac{\delta^2}{\delta \tilde{\Phi}^{\prime\prime}_{ia}(p) \delta \tilde{\Phi}^{\prime\prime}_{jb}(p) }S[\tilde{\Phi}]
\end{eqnarray}

We may represent
\begin{eqnarray}
{\cal P}^{\prime}_{(ia)(jb)} &=& \Omega_{ab}\delta^{ij} - G_{ab}\Upsilon^{ij} + \Pi^{\prime}_{(ia)(jb)},\label{P}\\ {\cal P}^{\prime\prime}_{(ia)(jb)} &=& \Omega_{ab}\delta^{ij} - G_{ab}\Upsilon^{ij} + B_{ab}\Upsilon^{ij} + \Pi^{\prime\prime}_{(ia)(jb)}\nonumber
\end{eqnarray}
where $\Pi$ is polarization operator. For its non - vanishing components we have ($a\ne i$):
\begin{widetext}
\begin{eqnarray}
&& \Pi^{\prime}_{(aa)(aa)}  \approx     -  \frac{2iN_c}{(2\pi)^4} \, \int  \frac{d^4 k }{(k^2 - m_a^2)((k+p)^2 - m_a^2)} \, \Big(k(p+k)+m_a^2\Big) \nonumber\\
&& \Pi^{\prime}_{(ia)(ia)}  \approx     -  \frac{2iN_c}{(2\pi)^4} \, \int  \frac{d^4 k }{(k^2 - m_i^2)((k+p)^2 - m_a^2)} \, k(p+k) \nonumber\\
&& \Pi^{\prime}_{(ia)(ai)}  \approx     -  \frac{2iN_c}{(2\pi)^4} \, \int  \frac{d^4 k }{(k^2 - m_i^2)((k+p)^2 - m_a^2)}\, m_im_a , \quad i\ne b \nonumber\\
&& \Pi^{\prime\prime}_{(aa)(aa)}  \approx     -  \frac{2iN_c}{(2\pi)^4} \, \int  \frac{d^4 k }{(k^2 - m_a^2)((k+p)^2 - m_a^2)} \, \Big(k(p+k)-m_a^2\Big) \nonumber\\
&& \Pi^{\prime\prime}_{(ia)(ia)}  \approx     -  \frac{2iN_c}{(2\pi)^4} \, \int  \frac{d^4 k }{(k^2 - m_i^2)((k+p)^2 - m_a^2)} \, k(p+k) \nonumber\\
&& \Pi^{\prime\prime}_{(ia)(ai)}  \approx     +  \frac{2iN_c}{(2\pi)^4} \, \int  \frac{d^4 k }{(k^2 - m_i^2)((k+p)^2 - m_a^2)}\, m_im_a, \quad i \ne b
\end{eqnarray}
\end{widetext}

\subsubsection{Calculation of polarization operator}

Let us introduce notations
\begin{eqnarray}
I(m) &=& \frac{i}{(2\pi)^4} \, \int d^4 l \, \frac{1}{l^2-
m^2}\label{Zeq}\\ &\approx& \frac{1}{16 \pi^2}(\Lambda^2-m^2\,{\rm log}\frac{\Lambda^2}{m^2})\nonumber\\
I(m_1,m_2,p) &=& - \frac{i}{(2\pi)^4} \, \int d^4 l \, \frac{1}{(l^2-
m_1^2)[(p-l)^2-m_2^2]}\nonumber
\end{eqnarray}

Using these notations we rewrite
\begin{eqnarray}
 \Pi^{\prime}_{(aa)(aa)} & \approx &     (-p^2+4m_a^2) N_c I(m_i,m_a,p) - 2N_c I(m_a) \nonumber\\
 \Pi^{\prime}_{(ia)(ia)} & \approx &   (-p^2+m_i^2+m_a^2) N_c I(m_i,m_a,p)\nonumber\\&& - N_c I(m_i) - N_c I(m_a) \nonumber\\
 \Pi^{\prime}_{(ia)(ai)} & \approx &    2 m_im_a N_c I(m_i,m_a,p)   \nonumber\\
 \Pi^{\prime\prime}_{(aa)(aa)} & \approx &    -p^2 N_c I(m_i,m_a,p) - 2N_c I(m_a) \nonumber\\
 \Pi^{\prime\prime}_{(ia)(ia)} & \approx &     (-p^2+m_i^2+m_a^2) N_c I(m_i,m_a,p) \nonumber\\&&- N_c I(m_i) - N_c I(m_a) \nonumber\\
 \Pi^{\prime\prime}_{(ia)(ai)} & \approx &    - 2 m_im_a N_c I(m_i,m_a,p)
\end{eqnarray}
At the same time the gap equation can be written as
\begin{equation}
{\omega}_a - {f}_a = 2 N_c I(m_a),
\end{equation}
for $a = t, \chi$.

\subsection{Evaluation of the scalar boson masses}

\subsubsection{Masses of charged scalar bosons}

Masses of charged bosons appear as the solutions of equation
\begin{equation}
{\rm Det}\, {\cal P}_{charged}(p^2) = 0
\end{equation}
where
\begin{widetext}
\begin{eqnarray}
{\cal P}_{charged}(p^2) = \left(\begin{array}{cc}
\begin{array}{c}(-p^2+m_t^2)\times\\\times N_c I(0,m_t,p)\\+f_t -N_c(I(m_t)-I(0)) \end{array}& \omega_{t\chi}  \\
\omega_{t\chi} &\begin{array}{c}(-p^2+m_\chi^2)\times\\\times N_c I(0,m_\chi,p)\\+f_\chi -N_c(I(m_\chi)-I(0))\end{array}\end{array}\right)
\label{mechargedexact}
\end{eqnarray}
\end{widetext}
Here parameters $\omega$ are the elements of matrix $\Omega$ in the basis of mass eigenstates and are given by Eq. (\ref{omegas}). Parameters $f$ are given by the next equation after Eq. (\ref{omegas}). In those equations $\alpha$ and $\theta$ are the mixing angles that enter the transformation from the basis of initial fermion fields to the mass eigenstates (see Eqs. (\ref{rot}), (\ref{rot2})). Integrals $I$ are defined in Eq. (\ref{Zeq}).

First of all, it is clear, that there is the massless charged scalar (one can check, that Eq. (\ref{mechargedexact})) has vanishing determinant at $p=0$. The second scalar is massive, and in order to evaluate its mass we are able to substitute $p^2 \approx m_\chi^2$ into Eq. (\ref{mechargedexact}).
Let us define the following quantities:
\begin{eqnarray}
N_c I(m_a,m_b,m_c) & = & Z^2_{abc}
\end{eqnarray}
Here
\begin{eqnarray}
&& N_c I(m_a,m_b,p)  =  \\ &&\frac{N_c}{16 \pi^2} \int_0^1 dx \, {\rm log}\,\frac{\Lambda^2}{m_a^2 x + m_b^2 (1-x) - p^2 x (1-x)}\nonumber
\end{eqnarray}
and we substitute $p^2 =m^2_c$. Notice that these integrals have imaginary parts for $m_c>m_a+m_b$, which correspond to the decays of the corresponding state with mass $m_c$ into the two fermions with masses $m_a$ and $m_b$. In the following we will chose the definition of logarithm (for negative values of arguments) in the above integral such that the imaginary part of the integral is positive. This will result in negative imaginary parts of the unstable scalar boson masses. If one of the arguments of $I(m_a,m_b,m_c)$ is zero, we denote the corresponding constant by $Z^2_{abc}$ with $a=0$, $b=0$, or $c=0$ correspondingly. In Euclidian region, where $p^2<0$ the integrals remain real - valued. Therefore, the mentioned imaginary parts do not affect stability of vacuum (to be considered after the Wick rotation). We also take into account that
\begin{eqnarray}
 Z^2_{ab0}=N_c I(m_a,m_b,0) & = & \frac{N_cI(m_b) - N_c I(m_a)}{m_a^2-m_b^2}
\end{eqnarray}
In Table \ref{table2} we represent real parts of $Z_{abc}^2$ for the example choices of arguments. These values should be compared to quantities
\begin{eqnarray}
  Z_t^2 & = & \frac{N_c}{16 \pi^2} {\rm log} \, \frac{\Lambda^2}{m_t^2}
  \nonumber\\
  Z_\chi^2 & = & \frac{N_c}{16 \pi^2} {\rm log} \, \frac{\Lambda^2}{m_\chi^2}
\end{eqnarray}
represented in Table \ref{table3}.
\begin{widetext}
\begin{table}
\begin{center}
\begin{small}
$\Lambda=10$ TeV, $m_\chi = 10 \, m_t$\\
\begin{tabular}{|c|c|c|c|c|c|c|c|c|c|c|c|c|c|c|c|c|}
\hline
& $m_3 = 0$ & $m_3 = m_t$ & $m_3 = m_H $ & $m_3 = m_\chi $ & $m_3 = 2 m_\chi $ \\
\hline
$\begin{array}{c} m_1=m_t \\ m_2 = 0  \end{array}$   & $0.1727103569$ & $0.1917080789$ & $0.1785398615$ & $0.1052842378$ & $0.07821589679$\\
\hline
$\begin{array}{c} m_1=m_2 =m_t   \end{array}$   & $0.1537126350$ & $0.1572500229$ & $0.1553811083$ & $0.1063659370$ & $0.07854932119$\\
\hline
$\begin{array}{c} m_1=m_t \\ m_2 = m_\chi  \end{array}$ & $0.08433889975$ & $0.08442804052$ & $0.08438340225$ & $0.09888674840$ & $0.08900924267$\\
\hline
$\begin{array}{c} m_1=m_\chi \\ m_2 = 0  \end{array}$ & $0.08522261432$ & $0.08531792115$ &  $0.08527018798$& $0.1042203362$ & $0.08856698817$\\
\hline
$\begin{array}{c} m_1=m_2=m_\chi   \end{array}$ &$0.06622489239$ & $0.06625658696$ & $0.06624073174$ & $0.06976228029$ & $0.1042203362$\\
\hline
\end{tabular}
\\\vspace{1mm}
$\Lambda=100$ TeV, $m_\chi = 10 \, m_t$\\
\begin{tabular}{|c|c|c|c|c|c|c|c|c|c|c|c|c|c|c|c|c|}
\hline
& $m_3 = 0$ & $m_3 = m_t$ & $m_3 = m_H $ & $m_3 = m_\chi $ & $m_3 = 2 m_\chi $ \\
\hline
$\begin{array}{c} m_1=m_t \\ m_2 = 0  \end{array}$   & $0.2601980996$ & $0.2791958215$ & $0.2660276041$ & $0.1927719804$ & $ 0.1657036394$\\
\hline
$\begin{array}{c} m_1=m_2=m_t  \end{array}$   & $0.2412003776$ & $0.2447377655$ & $0.2428688510$ & $0.1938536796$ & $0.1660370638$\\
\hline
$\begin{array}{c} m_1=m_t \\ m_2 = m_\chi  \end{array}$ & $0.1718266423$ & $0.1719157831$ & $0.1718711449$ & $0.1863744910$ & $0.1764969853$\\
\hline\
$\begin{array}{c} m_1=m_\chi \\ m_2 = 0  \end{array}$ & $0.1727103569$ & $0.1728056638$ &$0.1727579306$  & $0.1917080789$ & $0.1760547308$\\
\hline
$\begin{array}{c} m_1=m_2=m_\chi   \end{array}$ &$0.1537126350$ & $0.1537443296$ & $0.1537284743$ & $0.1572500229$ & $0.1917080789$\\
\hline
\end{tabular}
\\\vspace{1mm}
$\Lambda=100$ TeV, $m_\chi = 100 \, m_t$\\
\begin{tabular}{|c|c|c|c|c|c|c|c|c|c|c|c|c|c|c|c|c|}
\hline
& $m_3 = 0$ & $m_3 = m_t$ & $m_3 = m_H $ & $m_3 = m_\chi $ & $m_3 = 2 m_\chi $ \\
\hline
$\begin{array}{c} m_1=m_t \\ m_2 = 0  \end{array}$   & $0.2601980996$ & $0.2791958215$ & $0.2660276041$ & $0.1042397334$ &$0.07788940920$\\
\hline
$\begin{array}{c} m_1=m_2=m_t   \end{array}$   & $0.2412003776$ & $0.2447377655$ & $0.2428688510$ & $0.1042591342$& $0.07789491718$\\
\hline
$\begin{array}{c} m_1=m_t \\ m_2 = m_\chi  \end{array}$ & $0.08520511502$ & $0.08520605549$ & $0.08520558924$ & $0.1036341612$ & $0.08857432364$\\
\hline
$\begin{array}{c} m_1=m_\chi \\ m_2 = 0  \end{array}$ & $0.08522261432$ & $0.08522356424$ & $0.08522308927$ & $0.1042203362$ & $0.08856698817$\\
\hline
$\begin{array}{c} m_1=m_2=m_\chi   \end{array}$ & $0.06622489239$& $0.06622520902$ & $0.06622505070$  & $0.06976228029$ & $0.1042203362$\\
\hline
\end{tabular}
\\\vspace{1mm}
$\Lambda=1000$ TeV, $m_\chi = 100 \, m_t$\\
\begin{tabular}{|c|c|c|c|c|c|c|c|c|c|c|c|c|c|c|c|c|}
\hline
& $m_3 = 0$ & $m_3 = m_t$ & $m_3 = m_H $ & $m_3 = m_\chi $ & $m_3 = 2 m_\chi $ \\
\hline
$\begin{array}{c} m_1=m_t \\ m_2 = 0  \end{array}$   & $0.3476858422$ & $0.3666835641$ & $0.3535153468$ & $0.1917274761$ &$0.1653771518$\\
\hline
$\begin{array}{c} m_1=m_2=m_t   \end{array}$   & $0.3286881203$ & $0.3322255082$ & $0.3303565936$ & $0.1917468768$& $0.1653826598$\\
\hline
$\begin{array}{c} m_1=m_t \\ m_2 = m_\chi  \end{array}$ & $0.1726928576$ & $0.1726937981$ & $0.1726933318$ & $0.1911219038$ & $0.1760620662$\\
\hline
$\begin{array}{c} m_1=m_\chi \\ m_2 = 0  \end{array}$ & $0.1727103569$ & $0.1727113068$ & $0.1727108319$ & $0.1917080789$ & $0.1760547308$\\
\hline
$\begin{array}{c} m_1= m_2 = m_\chi  \end{array}$ &$0.1537126350$ & $0.1537129516$ & $0.1537127933$ & $0.1572500229$ & $0.1917080789$\\
\hline
\end{tabular}
\\\vspace{1mm}
$\Lambda=5\times 10^9$ TeV, $m_\chi = 100 \, m_t$\\
\begin{tabular}{|c|c|c|c|c|c|c|c|c|c|c|c|c|c|c|c|c|}
\hline
& $m_3 = 0$ & $m_3 = m_t$ & $m_3 = m_H $ & $m_3 = m_\chi $  & $m_3 = 2m_\chi $ \\
\hline
$\begin{array}{c} m_1=m_t \\ m_2 = 0  \end{array}$   & $0.9337636057$ & $0.9527613276$ & $0.9395931101$ & $0.7778052396$ & $0.7514549153$ \\
\hline
$\begin{array}{c} m_1=m_2=m_t   \end{array}$   & $0.9147658838$ & $0.9183032715$& $0.9164343571$ & $0.7778246403$ &$0.7514604233$\\
\hline
$\begin{array}{c} m_1=m_t \\ m_2 = m_\chi  \end{array}$ & $0.7587706210$ & $0.7587715618$ & $0.7587710286$ & $0.7771996673$ & $0.7621398296$\\
\hline
$\begin{array}{c} m_1=m_\chi \\ m_2 = 0  \end{array}$ & $0.7587881204$ & $0.7587890701$  &$0.7587885946$  & $0.7777858424$ & $0.7621324942$\\
\hline
$\begin{array}{c} m_1=m_2 = m_\chi    \end{array}$ & $0.7397903985$& $0.7397907150$ & $0.7397905568$ & $0.7433277863$ & $0.7777858424$\\
\hline
\end{tabular}
\end{small}
\caption{The values of ${\rm Re}\, Z_{a b c}^2$  for the values of parameters encountered in the text. Masses entering the corresponding integrals are denoted here by $m_a = m_1$, $m_b = m_2$, $m_c = m_3$. For $m_3>m_1+m_2$ the values of $Z_{a b c}^2$ have imaginary parts, which are omitted here.} \label{table2}
\end{center}
\end{table}
\end{widetext}

Let us assume, that the parameters $b$ and $g$ of the original Lagrangian are of the order of $m_\chi^2$. Then in order to calculate the second charged scalar boson mass (which is of the order of $m_\chi$) we may apply the approximation, in which the integrals $I(m_1,m_2,p)$  are substituted by $Z^2_{m_1 m_2 m_\chi}$. This approximation may be used at least for the rough evaluation of the scalar boson masses as follows from Tables \ref{table2} and \ref{table3}, i.e. its accuracy is within about $20$ per cents for $\Lambda = 10$ TeV, $m_\chi = 10 m_t$, and is improved, when the ratios $m_t/m_\chi$ and $m_\chi/\Lambda$ decrease. For example, for $\Lambda = 1000$ TeV, $m_t/m_\chi = 1/100$ the accuracy is within about five percents while for $\Lambda = 5\times 10^9$ TeV, $m_t/m_\chi = 1/100$ the accuracy is within two percents. Later we shall improve this accuracy substituting into the integrals $I(m_1,m_2,p)$ the values of $p^2$ equal to the calculated values of the corresponding scalar boson masses squared. Thus in the first approximation we come to
\begin{widetext}
\begin{eqnarray}
{\cal P}_{charged}(p^2) = \left(\begin{array}{cc}
\begin{array}{c}(-p^2+m_t^2)Z_{t0\chi}^2+f_t -m_t^2 Z_{t00}^2  \end{array}& \omega_{t\chi}  \\
\omega_{t\chi} &\begin{array}{c}(-p^2+m_\chi^2)Z^2_{ \chi 0 \chi}+f_\chi -m_\chi^2 Z^2_{\chi 0 0}\end{array}\end{array}\right)
\label{mechargedexact2}
\end{eqnarray}
\end{widetext}
Because of the $SU(2)_L$ symmetry of the original lagrangian we have $\omega^2_{t\chi}=f_tf_{\chi}$. Let us neglect the difference between $Z_{\chi0\chi}$ and $Z_{\chi 0 0}$. This gives for the channels that include the $b$ - quark
\begin{eqnarray}
&& M^{\prime (2)}_{H^{\pm}_t, H^{\pm}_{\chi}} = M^{\prime \prime (2)}_{H^{\pm}_t, H^{\pm}_{\chi}}  = 0\label{charged}\\
&&\Big[M^{\prime (1)}_{H^{\pm}_t, H^{\pm}_{\chi}}\Big]^2 = \frac{1}{2}(\frac{g_\chi}{Z_{\chi 0\chi}^2}(1+w^2\gamma_\chi^2)+m_\chi^2\delta_\chi)\nonumber\\&&+\frac{1}{2}\sqrt{(\frac{g_\chi}{Z_{\chi 0\chi}^2}(1+w^2\gamma_\chi^2)-m_\chi^2\delta_\chi)^2+4m_\chi^2\delta_\chi \frac{g_\chi}{Z^2_{\chi 0 \chi}}}
\nonumber\\&&\approx  \frac{g_\chi}{Z_{\chi 0\chi}^2}(1+w^2\gamma_\chi^2) + m_\chi^2 \delta_\chi \frac{1}{1+w^2\gamma_\chi^2},\nonumber\\ &&
\gamma_\chi = \frac{Z_{\chi 0\chi}}{Z_{t 0 \chi}}, \quad \delta_\chi = \frac{Z_{\chi 0\chi}^2 - Z_{\chi 0 0}^2}{Z^2_{\chi 0 \chi}}\nonumber
\end{eqnarray}
At it was mentioned above, in this channel the charged exactly massless Goldstone boson appears (to be eaten by
the $W$ - boson) that corresponds to the spontaneous breakdown of $SU(2)_L$. Notice, that constant $Z^2_{t0\chi}$ has an imaginary part because we consider the case $m_\chi > m_t$. As a result $M^{(1)}_{H^{\pm}_t, H^{\pm}_{\chi}}$ receives imaginary part as well, which corresponds to the decay of the charged scalar field into the pair $\bar{t}b$ (or $\bar{b}t$).
As it was mentioned above, in order to improve the estimate of this mass, we should substitute into Eq. (\ref{charged})  constants $N_c I(m_t,0, M^{\prime (1)}_{H^{\pm}_t, H^{\pm}_{\chi}})$ and $N_c I( m_\chi, 0, M^{\prime (1)}_{H^{\pm}_t, H^{\pm}_{\chi}})$ instead of $Z^2_{t 0 \chi}$ and $Z^2_{\chi 0 \chi}$ with the masses $M^{\prime (1)}_{H^{\pm}_t, H^{\pm}_{\chi}}$ evaluated using the first order approximation of the above expression.

\subsubsection{Masses of CP - odd neutral scalar bosons}
For the CP - odd neutral states we use the
basis
$A_t=\tilde{\Phi}^{\prime\prime}_{tt} \sim [\bar{t}_L t_R - \bar{t}_R t_L]$, $A_{\chi}=\tilde{\Phi}^{\prime\prime}_{t\chi} \sim[\bar{t}_L \chi_R - \bar{\chi}_Rt_L]$,
$\pi_t = \tilde{\Phi}^{\prime\prime}_{\chi t} \sim [\bar{\chi}_L t_R - \bar{t}_R \chi_L]$,
  $\pi_{\chi}=\tilde{\Phi}^{\prime\prime}_{\chi \chi} \sim [\bar{\chi}_L \chi_R - \bar{\chi}_R
\chi_L]$.
We should solve equation
\begin{equation}
{\rm Det}\, {\cal P}^{\prime\prime}_{}(p^2) = 0
\end{equation}
The matrix function ${\cal P}^{\prime\prime}_{}(p^2)$ in the above mentioned basis is given by:
\begin{widetext}
\begin{eqnarray}
\left(\begin{array}{cccc}
\begin{array}{c}(-p^2) N_c I(m_t,m_t,p)\\+f_t - (g_t-b_t) \lambda_t\end{array}& \omega_{t\chi} - (g_{t\chi}-b_{t\chi}) \lambda_t&  -(g_t-b_t) \lambda_{t\chi} &-(g_{t\chi}-b_{t\chi})\lambda_{t\chi} \\
\omega_{t\chi} - (g_{t\chi}-b_{t\chi})\lambda_t&\begin{array}{c}(-p^2+ m_{t}^2+m_\chi^2)\times\\\times N_cI(m_t,m_\chi,p)\\ + N_c(I(m_\chi)-I(m_t))\\+ f_{\chi}- (g_\chi-b_\chi) \lambda_t\end{array}&\begin{array}{c}- 2 m_{t} m_{\chi}\times\\\times N_cI(m_t,m_\chi,p)\\ - (g_{t\chi}-b_{t\chi}) \lambda_{t\chi}\end{array} & -(g_\chi-b_\chi)\lambda_{t\chi} \\
-(g_t-b_t) \lambda_{t\chi} & \begin{array}{c} -2 m_{t} m_{\chi}
 N_cI(m_t,m_\chi,p) \\ - (g_{t\chi}-b_{t\chi})\lambda_{t\chi}\end{array} &\begin{array}{c}(-p^2+ m_{t}^2+m_\chi^2)\times\\\times N_cI(m_t,m_\chi,p)\\ - N_c(I(m_\chi)-I(m_t))\\+ f_t- (g_t-b_t)\lambda_{\chi}\end{array} &\omega_{t\chi}-(g_{t\chi}-b_{t\chi})\lambda_{\chi}
\\
-(g_{t\chi}-b_{t\chi})\lambda_{t\chi}&-(g_\chi-b_\chi)\lambda_{t\chi} & \omega_{t\chi}-(g_{t\chi}-b_{t\chi})\lambda_{\chi} & \begin{array}{c}(-p^2) N_c I(m_\chi,m_\chi,p)\\+f_{\chi}-(g_\chi-b_\chi) \lambda_{\chi}\end{array}
 \end{array}\right)
\label{moddexact}
\end{eqnarray}
\end{widetext}
Here parameters $\lambda$ are given by
\begin{equation}
\lambda_t = {\rm sin}^2 \theta , \quad \lambda_{t\chi} = {\rm sin}\,\theta \, {\rm cos}\,\theta, \quad \lambda_{\chi}= {\rm cos}^2\theta\label{lambdas}
\end{equation}
 Parameters $g$ are the elements of matrix $G$ in the basis of mass eigenstates and are given by Eq. (\ref{G3A}). Parameters $b$ are the elements of matrix $B$ in the same basis. Parameters $\omega$ are the elements of matrix $\Omega$ in the basis of mass eigenstates and are given by Eq. (\ref{omegas}). Parameters $f$ are given by the next equation after Eq. (\ref{omegas}).In those equations $\alpha$ and $\theta$ are the mixing angles that enter the transformation from the basis of initial fermion fields to the mass eigenstates (see Eqs. (\ref{rot}), (\ref{rot2})). Integrals $I$ are defined in Eq. (\ref{Zeq}).

First of all, we have checked using MAPLE package, that the determinant of Eq. (\ref{moddexact}) for $p=0$ is zero, which means, that there exists the CP odd neutral Goldstone boson to be eaten by the Z boson.
Again, we assume, that parameters $b$ and $g$ are of the order of $m_\chi^2$. Therefore, the remaining masses are of the order of $m_\chi$. And as for the charged scalar bosons we first apply the approximation, in which all integrals $I$ are substituted by the factors  $Z^2_{m_1 m_2 m_\chi}$.

Next, we neglect the ratio $m_t/m\chi$ and arrive at the following expression for
${\cal P}^{\prime\prime}(p^2)$:
\begin{widetext}
\begin{eqnarray}
 \left(\begin{array}{cccc}
-p^2 Z^2_{tt\chi} + \frac{g_{t\chi}^2}{g_{\chi}} & g_{t\chi}  & 0 &0 \\
g_{t\chi} & (- p^2 + m_\chi^2) Z^2_{t\chi \chi} - m_\chi^2 Z^2_{t\chi 0} + g_{\chi}& 0& 0 \\
0 & 0  & (-p^2 + m_{\chi}^2) Z_{t\chi \chi}^2 +  m_{\chi}^2 Z_{t\chi 0 }^2 + \frac{g_{t\chi}^2}{g_{\chi}} - g_t + b_t & b_{t\chi}\\
0&0 &b_{t\chi} & -p^2 Z_{\chi\chi \chi}^2 + b_\chi
 \end{array}\right) \nonumber
\end{eqnarray}
\end{widetext}

The exactly massless Goldstone boson to be eaten by the Z - boson is mostly given the combination of $A_t$ and $A_{\chi}$. The masses of the remaining CP - odd neutral scalar bosons in this approximation are
\begin{eqnarray}
 M^{(1)}_{A_t A_{\chi}}&=&0, \nonumber\\ \Big[M^{(2)}_{A_t A_{\chi}}\Big]^2& =& \frac{1}{2}(\frac{g_\chi}{Z_{t\chi \chi}^2}(1+w^2\gamma_t^2)+m_\chi^2\delta_t)\nonumber\\&&+\frac{1}{2}\sqrt{(\frac{g_\chi}{Z_{t\chi \chi}^2}(1+w^2\gamma_t^2)-m_\chi^2\delta_t)^2+4m_\chi^2\delta_t \frac{g_\chi}{Z^2_{t\chi\chi}}}
\nonumber\\&&\approx  \frac{g_\chi}{Z_{t\chi \chi}^2}(1+w^2\gamma_t^2) + m_\chi^2 \delta_t \frac{1}{1+w^2\gamma_t^2},\nonumber\\ &&
\gamma_t = \frac{Z_{t\chi \chi}}{Z_{tt\chi}}, \quad \delta_t = \frac{Z_{t\chi \chi}^2 - Z_{t\chi 0}^2}{Z^2_{t\chi\chi}} \nonumber\\
 M^{(1,2)}_{\pi_{\chi}, \pi_t} &=&\Big(m_{\chi}^2+\frac{b_{\chi}+\tilde{b}_t}{2Z_{t\chi\chi}^2}\pm \Big[\Big(m_{\chi}^2+\frac{b_{\chi}+\tilde{b}_t}{2Z_{t\chi\chi}^2}\Big)^2 \nonumber\\&&- \frac{b_{\chi}\tilde{b}_t}{Z_{t\chi\chi}^4}- 2 m^2_{\chi}\frac{b_{\chi}}{Z_{t\chi\chi}^2}+\frac{b_{t\chi}^2}{Z_{t\chi\chi}^4}\Big]^{1/2}\Big)^{1/2},\nonumber
\end{eqnarray}
where
\begin{equation}
\tilde{b}_t = b_t - g_t + \frac{g_{t\chi}^2}{g_{\chi}}
\end{equation}
In expression for $M^{(1,2)}_{\pi_{\chi}, \pi_t}$ we neglect the difference between $Z_{\chi\chi\chi}$, $Z_{t\chi 0}$, and $Z_{t\chi\chi}$ for simplicity. In practical calculation of these masses for the particular example choices of parameters (see below Sect. \ref{SectPart}) we take into account this difference. It appears, that the above expression is only the first approximation, and the actual values of masses may have imaginary parts, which correspond to the decays of the given states to the pairs of fermions (see Sect. \ref{SectPart}, where we substitute into the mass matrix constants $N_c I(m_\chi,m_\chi, M^{(1,2)}_{\pi_{\chi}, \pi_t})$ and $N_c I(m_t, m_\chi, M^{(1,2)}_{\pi_{\chi}, \pi_t})$ instead of $Z^2_{\chi\chi\chi}$ and $Z^2_{t\chi\chi}$ with the masses $M^{(1,2)}_{\pi_{\chi}, \pi_t}$ evaluated using the first order approximation of the above expression).  Notice, that $Z^2_{t t \chi}$ itself has nonzero imaginary part from the very beginning because $m_\chi > 2 m_t$. Therefore, the mass $M^{(2)}_{A_t A_{\chi}}$ has imaginary part, which also means that the corresponding state is unstable and is ably to decay into the pair $\bar{t}t$.

\subsubsection{Masses of CP - even neutral scalar bosons}

For the CP - even neutral states we use the
basis
$h_t = \tilde{\Phi}^{\prime}_{tt} \sim [\bar{t}_L t_R + \bar{t}_R t_L]$, $h_{\chi} = \tilde{\Phi}^{\prime}_{t\chi} \sim[\bar{t}_L \chi_R + \bar{\chi}_Rt_L]$,
$\varphi_t = \tilde{\Phi}^{\prime}_{\chi t} \sim [\bar{\chi}_L t_R + \bar{t}_R \chi_L]$,
  $\varphi_{\chi}=\tilde{\Phi}^{\prime}_{\chi \chi} \sim [\bar{\chi}_L \chi_R + \bar{\chi}_R
\chi_L]$.
In order to calculate the scalar boson masses  we need to solve equation
\begin{equation}
{\rm Det}\, {\cal P}^{\prime}_{}(p^2) = 0
\end{equation}
 and to identify the lowest solution of this equation with $M_H^2$. The matrix function ${\cal P}^{\prime}_{}(p^2)$ is
\begin{widetext}
\begin{eqnarray}
\left(\begin{array}{cccc}
\begin{array}{c}(-p^2+4m_t^2)\times\\\times N_c I(m_t,m_t,p)\\+f_t - g_t \lambda_t\end{array}& \omega_{t\chi} - g_{t\chi} \lambda_t&  -g_t \lambda_{t\chi} &-g_{t\chi}\lambda_{t\chi} \\
\omega_{t\chi} - g_{t\chi}\lambda_t&\begin{array}{c}(-p^2+ m_{t}^2+m_\chi^2)\times\\\times N_cI(m_t,m_\chi,p)\\ + N_c(I(m_\chi)-I(m_t))\\+ f_{\chi}- g_{\chi} \lambda_t\end{array}&\begin{array}{c} 2 m_{t} m_{\chi}\times\\\times N_cI(m_t,m_\chi,p)\\ - g_{t\chi} \lambda_{t\chi}\end{array} & -g_{\chi}\lambda_{t\chi} \\
-g_t \lambda_{t\chi} & \begin{array}{c} 2 m_{t} m_{\chi}
\times\\\times N_cI(m_t,m_\chi,p)- g_{t\chi}\lambda_{t\chi}\end{array} &\begin{array}{c}(-p^2+ m_{t}^2+m_\chi^2)\times\\\times N_cI(m_t,m_\chi,p)\\ - N_c(I(m_\chi)-I(m_t))\\+ f_t- g_t\lambda_{\chi}\end{array} &\omega_{t\chi}-g_{t\chi}\lambda_{\chi}
\\
-g_{t\chi}\lambda_{t\chi}&-g_{\chi}\lambda_{t\chi} & \omega_{t\chi}-g_{t\chi}\lambda_{\chi} & \begin{array}{c}(-p^2+4m_\chi^2)\times\\\times N_c I(m_\chi,m_\chi,p)\\+f_{\chi}-g_{\chi} \lambda_{\chi}\end{array}
 \end{array}\right)
\label{mevenexact}
\end{eqnarray}
\end{widetext}
Here parameters $\lambda$ are given by Eq. (\ref{lambdas}), parameters $g$ are the elements of matrix $G$ in the basis of mass eigenstates and are given by Eq. (\ref{G3A}). Parameters $\omega$ are the elements of matrix $\Omega$ in the basis of mass eigenstates and are given by Eq. (\ref{omegas}). Parameters $f$ are given by the next equation after Eq. (\ref{omegas}).In those equations $\alpha$ and $\theta$ are the mixing angles that enter the transformation from the basis of initial fermion fields to the mass eigenstates (see Eqs. (\ref{rot}), (\ref{rot2})). Integrals $I$ are defined in Eq. (\ref{Zeq}).

Our aim is to check that there exists the region of parameters, where the lowest CP - even neutral scalar boson mass is given by $M_H \approx m_t/\sqrt{2}$. One can easily  find, that in the zero order approximation in powers of $m_t$ we have $M_H^{(0)}=0$. In order to calculate the first and the second order approximations we substitute $p^2 = M_H^2 = m_t^2/2$ into the integrals $I(m_1,m_2,p)$ in Eq. (\ref{mevenexact}). Since we know the exact value of the required mass, we can do this in order to evaluate the region of parameters, which gives the correct lightest Higgs boson mass. For the calculation of this lightest CP even scalar boson mass we use the more refined approximation than for the calculation of the other scalar boson masses. Namely, in order to calculate the correction to $[M_H^{(0)}]^2 = 0$ proportional to $m_t^2$ we consider first the zero order approximation to ${\cal P}^{\prime}_{}(p^2)$ (with $p^2 = M_H^2$ substituted into the integrals $I(m_1,m_2,p)$) in the form
\begin{widetext}
\begin{equation}
 \left(\begin{array}{cccc}
-p^2 Z^2_{ttH} + \frac{g_{t\chi}^2}{g_{\chi}} & g_{t\chi}  & 0 &0 \\
g_{t\chi} & (- p^2 + m_\chi^2) Z^2_{t\chi H} - m_\chi^2 Z^2_{t\chi 0} + g_{\chi}& 0& 0 \\
0 & 0  & (-p^2 + m_{\chi}^2) Z_{t\chi H}^2 +  m_{\chi}^2 Z_{t\chi 0}^2 + \frac{g_{t\chi}^2}{g_{\chi}} - g_t & 0
\\
0&0 &0 & (-p^2+4 m_{\chi}^2)Z_{\chi\chi H}^2
 \end{array}\right) \nonumber
 \end{equation}
 \end{widetext}

The zero order in the powers of $m_t$ gives the following value of the smallest mass:
\begin{widetext}
\begin{eqnarray}
[M_H^{(0)}]^2 &=& \frac{1}{2}(\frac{g_\chi}{Z_{t\chi H}^2}(1+w^2\gamma^2)+m_\chi^2\delta)-\frac{1}{2}\sqrt{(\frac{g_\chi}{Z_{t\chi H}^2}(1+w^2\gamma^2)-m_\chi^2\delta)^2+4m_\chi^2\delta \frac{g_\chi}{Z^2_{t\chi H}}}
\nonumber\\&\approx &  m_\chi^2 \delta \frac{w^2\gamma^2}{1+w^2\gamma^2},\nonumber\\ &&
\gamma = \frac{Z_{t\chi H}}{Z_{t t H}}, \quad \delta = \frac{Z_{t\chi H}^2 - Z_{t \chi 0}^2}{Z^2_{t\chi H}}, \quad w=\frac{g_{t\chi}}{g_{\chi}}
\end{eqnarray}
\end{widetext}
and the corresponding Higgs scalar field
\begin{eqnarray}
H & \approx & \sqrt{2}Z_{ttH} \frac{h_{t} - \omega \gamma \zeta h_{\chi}}{\sqrt{1+
w^2 \gamma^2 \zeta^2}},\nonumber\\&&
\zeta=1-\frac{m_\chi^2}{g_\chi(1+w^2\gamma^2)}\delta \label{Hconst0}
\end{eqnarray}
(The kinetic term for this field is normalized in such a way, that it is given by $\frac{1}{2} H^2\hat{p}^2 H$).

We take into account, that $\delta \ll 1$, i.e. that the difference between $Z^2_{t\chi H}$ and $Z^2_{t\chi 0} $ is small. For example, for  $\Lambda = 1000$ TeV, $m_\chi = 100 m_t$ we have $\delta \sim 3 \times 10^{-6}$ as follows from Table \ref{table2}. Thus, this is a reasonable approximation that allows to evaluate the lightest mass even in the presence of a fine tuning. In order to calculate the corrections to the value of $M_H$ proportional to $m_t^2$ we use the ordinary second order perturbation theory applied to the lowest eigenvalue of the following matrix $\hat{M}_{\rm even}^2$ (calculated up to the terms $\sim m_t^2$):
\begin{widetext}
\begin{eqnarray}
\frac{1}{Z_{t\chi H}^2}\left(\begin{array}{cccc}
\begin{array}{c}\frac{g_{t\chi}^2}{g_{\chi}}\frac{Z_{t\chi H}^2}{Z_{ttH}^2} + (4 Z_{t\chi H}^2 m_{\chi}^2\\ + [g_t w^2 - 2 \, g_{\chi} w^4]\frac{Z_{t\chi H}^2}{Z_{ttH}^2})\frac{m_t^2}{m^2_{\chi}} \end{array}& \Big[g_{\chi}w + w (g_t - 2 w^2 g_{\chi})\frac{m_t^2}{m_{\chi}^2}\Big]\frac{Z_{t\chi H}}{Z_{ttH}}  & \Big[-g_t w \frac{m_t}{m_{\chi}}\Big]\frac{Z_{t\chi H}}{Z_{ttH}}  &- w^2 g_{\chi} \frac{m_t}{m_{\chi}}\frac{Z_{t\chi H}}{Z_{ttH}}\frac{Z_{t\chi H}}{Z_{\chi\chi H}}\\
\begin{array}{c}\Big[ w (g_t - 2 w^2 g_{\chi})\frac{m_t^2}{m_{\chi}^2}\\ +g_{\chi}w \Big]\frac{Z_{t\chi H}}{Z_{ttH}} \end{array}&\begin{array}{c} g_{\chi}+m^2_\chi(Z_{t\chi H}^2 - Z_{t\chi 0}^2)\\+\Big((Z_{t\chi H}^2 +Z_{t\chi 0}^2) m^2_{\chi}\\-g_{\chi}w^2\Big)\frac{m_t^2}{m_{\chi}^2} \end{array}& (2 Z_{t\chi H}^2 m_{\chi}^2-w^2g_{\chi})\frac{m_t}{m_{\chi}} & -w g_{\chi}\frac{m_t}{m_{\chi}}\frac{Z_{t\chi H}}{Z_{\chi\chi H}} \\
-g_t w \frac{m_t}{m_{\chi}}\frac{Z_{t\chi H}}{Z_{ttH}} & (2 Z_{t\chi H}^2 m_{\chi}^2-w^2g_{\chi})\frac{m_t}{m_{\chi}}   & \begin{array}{c}\tilde{g}_t + (Z_{t\chi H}^2 - Z_{t\chi 0}^2)m_{t}^2\\+(3 g_t - 2 g_{\chi}w^2)w^2\frac{m_t^2}{m_{\chi}^2}\end{array}& g_t w \frac{m_t}{m_{\chi}}\frac{Z_{t\chi H}}{Z_{\chi\chi H}}
\\
- w^2 g_{\chi} \frac{m_t}{m_{\chi}} \frac{Z_{t\chi H}}{Z_{ttH}}\frac{Z_{t\chi H}}{Z_{\chi\chi H}}&-w g_{\chi}\frac{m_t}{m_{\chi}} \frac{Z_{t\chi H}}{Z_{\chi\chi H}} &g_t w \frac{m_t}{m_{\chi}} \frac{Z_{t\chi H}}{Z_{\chi\chi H}}& \begin{array}{c}4 Z_{t\chi H}^2 m_{\chi}^2\\+g_{\chi}w^2\frac{m_t^2}{m_{\chi}^2}\frac{Z^2_{t\chi H}}{Z^2_{\chi\chi H}}\end{array}
 \end{array}\right)\nonumber
\end{eqnarray}
\end{widetext}
Here $\tilde{g}_t= (Z_{t\chi H}^2 + Z_{t\chi 0}^2)m_{\chi}^2+w^2 g_{\chi}-g_t$ while $w=\frac{g_{t\chi}}{g_{\chi}}$.
This mass matrix is defined in the basis $\tilde{\tilde{\Phi}}^\prime =  (Z_{ttH} h_t, Z_{t\chi H} h_\chi, Z_{t\chi H} \phi_t, Z_{\chi\chi H} \phi_\chi)^T$, in which the effective action for $p^2$ around $m_t^2/2$ has the form
\begin{equation}
S_{even}\approx \int \frac{d^4 p}{(2\pi^4)} \Big[\tilde{\tilde{\Phi}}^\prime\Big]^T (\hat{p}^2-\hat{M}_{\rm even}^2)\tilde{\tilde{\Phi}}^\prime
\end{equation}

In the correction to $M_H^2$ proportional to $m_t^2$ we may neglect $\delta$. The resulting expression for $M_H^2$ has the form:
\begin{widetext}
\begin{eqnarray}
M^2_H\approx   m_\chi^2 \frac{Z_{t\chi H}^2 - Z_{t \chi 0}^2}{Z_{t\chi H}^2} \frac{w^2\frac{Z^2_{t\chi H}}{Z_{ttH}^2}}{1+w^2\frac{Z_{t\chi H}^2}{Z_{ttH}^2}} + 4 m^2_{t} \, \frac{1-w^2\frac{Z_{t\chi H}^2}{Z_{ttH}^2}\Big(\frac{Z_{t\chi H}^2 \Big[1+\frac{Z^2_{t\chi 0}}{Z_{t\chi H}^2}\Big]^2 m^2_{\chi}}{\tilde{g}_t} - \Big[1+\frac{Z^2_{t\chi 0}}{Z_{t\chi H}^2}\Big]\Big)}{1+w^2\frac{Z_{t\chi H}^2}{Z_{ttH}^2}} + O(m_t^4) \label{mha}
\end{eqnarray}
\end{widetext}

In the following we may neglect $\delta$ in all other expressions. This means, in particular, that $\zeta = 1$ in Eq. (\ref{Hconst0}). Notice, that Eq. (\ref{mha}) is valid only for the small values of ratio $m_t/m_\chi$. Our numerical analysis demonstrates, that Eq. (\ref{mha}) gives accuracy witin one percent for the calculation of the lightest neutral Higgs boson mass for $\Lambda = 1000$ TeV and $m_t/m_\chi = 1/100$, while for $\Lambda = 10$ TeV and $m_t/m_\chi = 1/10$ it gives the accuracy of about $10$ percent.

In order to calculate the remaining masses (that are of the order of $m_\chi$) we neglect the ratio $m_t/m_{\chi}$, and consider
${\cal P}^{\prime}(p^2)$ in the form
\begin{widetext}
\begin{equation}
 \left(\begin{array}{cccc}
-p^2 Z^2_{tt\chi} + \frac{g_{t\chi}^2}{g_{\chi}} & g_{t\chi}  & 0 &0 \\
g_{t\chi} & (- p^2 + m_\chi^2) Z^2_{t\chi \chi} - m_\chi^2 Z^2_{t\chi 0} + g_{\chi}& 0& 0 \\
0 & 0  & (-p^2 + m_{\chi}^2) Z_{t\chi \chi}^2 +  m_{\chi}^2 Z_{t\chi 0}^2 + \frac{g_{t\chi}^2}{g_{\chi}} - g_t & 0
\\
0&0 &0 & (-p^2+4 m_{\chi}^2)Z_{\chi\chi \chi}^2
 \end{array}\right)\nonumber
 \end{equation}
\end{widetext}

This gives
\begin{eqnarray}
\Big[ M^{(2)}_{h_th_{\chi}}\Big]^2&=&
\label{evens}  \frac{1}{2}(\frac{g_\chi}{Z_{t\chi \chi}^2}(1+w^2\gamma_t^2)+m_\chi^2\delta_t)\nonumber\\&&+\frac{1}{2}\sqrt{(\frac{g_\chi}{Z_{t\chi \chi}^2}(1+w^2\gamma_t^2)-m_\chi^2\delta_t)^2+4m_\chi^2\delta_t \frac{g_\chi}{Z^2_{t\chi\chi}}}
\nonumber\\&&\approx  \frac{g_\chi}{Z_{t\chi \chi}^2}(1+w^2\gamma_t^2)+ m_\chi^2 \delta_t \frac{1}{1+w^2\gamma_t^2},\nonumber\\ &&
\gamma_t = \frac{Z_{t\chi \chi}}{Z_{tt\chi}}, \quad \delta_t = \frac{Z_{t\chi \chi}^2 - Z_{t\chi 0}^2}{Z^2_{t\chi\chi}}\nonumber\\
 M_{\varphi_{\chi}} &\approx &2 m_{\chi}\nonumber\\
  M_{\varphi_t} &\approx &\frac{\sqrt{(Z_{t\chi \chi}^2 + Z_{t\chi 0}^2)m_{\chi}^2+w^2 g_{\chi}-g_t}}{Z_{t\chi \chi}}
\end{eqnarray}
Recall, that $Z^2_{t t \chi}$ has nonzero imaginary part because $m_\chi > 2 m_t$. Therefore, the mass $M^{(2)}_{h_th_{\chi}}$ has imaginary part, which means that the corresponding state is unstable and may decay into the pair $\bar{t}t$. Again, as for the CP - odd states the above expression for $M_{\varphi_{t}}$ is only the first approximation. It actually may have an imaginary part, which results from the more precise estimate
\begin{eqnarray}
M_{\varphi_t} &\approx &\frac{\sqrt{(Z_{t\chi \varphi_t}^2 + Z_{t\chi 0}^2)m_{\chi}^2+w^2 g_{\chi}-g_t}}{Z_{t\chi \varphi_t}}
\end{eqnarray}
We should substitute here $Z^2_{m_t m_\chi \varphi_t} = N_c I(m_t,m_\chi,M_{\varphi_t})$ with the first order approximation for $M_{\varphi_t}$. If the latter mass is larger, than the sum of $m_t$ and $m_\chi$, the value of $M_{\varphi_t}$ acquires imaginary part. In practical calculations in Sect. \ref{SectPart} we apply the same procedure to all other composite scalar boson masses.

\begin{widetext}
\begin{table}
\begin{center}
\begin{small}
\begin{tabular}{|c|c|c|c|c|c|c|c|c|c|c|c|c|c|c|c|c|}
\hline
&  $Z_\chi$ & $Z_t$ \\
\hline
$\begin{array}{c}\Lambda=10 \, {\rm  TeV}\\m_\chi = 10 \, m_t \end{array}$  & $0.06622489236$ & $0.1537126349$ \\
\hline
$\begin{array}{c}\Lambda=100 \, {\rm  TeV}\\m_\chi = 10 \, m_t \end{array}$  & $0.1537126349$ & $0.2412003776$ \\
\hline
$\begin{array}{c}\Lambda=100 \, {\rm  TeV}\\m_\chi = 100 \, m_t \end{array}$ & $0.06622489236$ & $0.2412003776$ \\
\hline
$\begin{array}{c}\Lambda=1000 \, {\rm  TeV}\\m_\chi = 100 \, m_t \end{array}$ & $0.1537126349$ & $0.3286881202$ \\
\hline
$\begin{array}{c}\Lambda=5\times 10^9 \, {\rm  TeV}\\m_\chi = 100 \, m_t \end{array}$ & $0.7397903985$ & $0.9147658838$ \\
\hline
\end{tabular}
\end{small}
\caption{The values of  $Z_t^2$ and $Z_\chi^2$ for certain values of parameters.} \label{table3}
\end{center}
\end{table}
\end{widetext}

\subsection{Phenomenology}

\label{SectPhenomenology}

\subsubsection{PNG candidate for the $125$ GeV Higgs}

Symmetry breaking pattern in the given model is as follows. Without the $SU(3)$ breaking terms we have the original global $SU(3)_L \otimes U(1)_L\otimes U(1)_{t, R}\otimes U(1)_{\chi, R}$ symmetry that is broken spontaneously down to $U(1)_{t}\otimes U(1)_{\chi}\otimes U(1)_b$. (Here $U(1)_{t}$, $U(1)_{\chi}$ act on the left and the right - handed components of $t$ and $\chi$  while $U(1)_b$ acts on the left - handed b - quark.) As a result among the $12$ components of $\tilde{\Phi}$ we have $8$ Goldstone bosons. There are $4$ massless states that are composed of $b$ - quark: $H_t^{\pm}, H_{\chi}^{\pm}$, there are $3$ CP - odd massless states $A_t, \pi_{\chi}$ and $\frac{A_{\chi}m_{\chi} + \pi_t m_{t}}{\sqrt{m_t^2+m_{\chi}^2}}$, and there is one CP - even massless state $\frac{m_{\chi} h_{\chi} - m_{t} \varphi_t}{\sqrt{m_t^2 +
m_{\chi}^2}}$.

When the $SU(3)$ breaking modification of the model is turned on, the original symmetry is reduced to $SU(2)_L \otimes U(1)_L$. This symmetry is broken spontaneously down to $U(1)_b$. As a result we have $3$ exactly massless Goldstone bosons to be eaten by $W^{\pm}$ and $Z$, and $5$ Pseudo - Goldstone bosons.
When the $SU(3)$ breaking terms are turned on, the structure of the scalar spectrum is changed.

 We consider the particular case, when there are the following relations between the parameters of the model:
\begin{eqnarray}
&&m_t^2 \ll  g_{t,\chi,t\chi} \sim m_\chi^2 \ll \omega_t\sim\omega_{\chi} \sim \Lambda^2 \label{partcase}
\end{eqnarray}
In the considered case
the lightest CP - even state $H$ is given mostly by the combination of $h_t,h_{\chi}$ instead of the combination of $\varphi_t,h_{\chi}$ (Eq. (\ref{Hconst0})).
This state realizes the conventional top quark condensation scenario, when $g_{t\chi} \ll g_{\chi}$ so that it is composed mostly of $\bar{t}t$.  When $m_t = 0$ it becomes massless. The presence of nonzero $m_t$ gives it the mass. The expression for the mass in general case is very complicated. It depends on 5 parameters: $g_t, g_{\chi}, g_{t\chi}, m_t, m_{\chi}$. The leading order in $m_t$ is $M_H^2 \sim m_t^2$. We demonstrate, that there exists the appropriate choice of the remaining parameters such that the Higgs boson mass is set to its observed value that is $M_H^2 \approx \frac{m_t^2}{2}$.

Above we derived Eq. (\ref{mha}) for the Higgs boson mass, which is valid at $m_t \ll m_\chi$. Parameters $g$ entering this expression are the elements of matrix $G$ in the basis of mass eigenstates and are given by Eq. (\ref{G3A}).
The corresponding values of parameters satisfy relation $M_H = m_t/\sqrt{2}$, and $g_t, g_\chi, g_{t\chi},Z, m_t, m_\chi$ are expressed through the mentioned above bare parameters via the gap equations  Eq. (\ref{gapeqs}), and through Eq. (\ref{Zeq}), and  Eqs. (\ref{G3A}) and (\ref{theta}) that allow to determine precisely $\theta$ and $\alpha$ as functions of $g^{(0)}_{t,\chi,t\chi}$ and then $g_{t,\chi,t\chi}$ as functions of $g^{(0)}_{t,\chi,t\chi}$. (As it was already mentioned, the corresponding expressions are so complicated that we do not represent them here.)

In Euclidean space the effective potential for the CP even neutral scalar bosons and charged scalar bosons is stable if
\begin{equation}
g_{\chi}>0, \quad \tilde{g}_t>0 \label{stability}
\end{equation}
The appropriate choice of parameters $b_t,b_{\chi}, b_{t\chi}$ always allows to make stable the effective potential for the  CP odd scalar bosons (those parameters do not enter Eq.(\ref{mha})). Therefore, we consider Eq. (\ref{stability}) as the condition for the stability of vacuum.

\subsubsection{Electroweak symmetry breaking}

Above we calculated effective action for the field $\tilde{\Phi}$, which is the fluctuation above the condensate. We may consider the part of this effective action that contains $\hat{p}^2$ and
reconstruct the whole effective action for the field $\Phi$:
\begin{widetext}
\begin{eqnarray}
S &\approx&\int d^4 x \left(\begin{array}{c}{\Phi}_{bt}\\ {\Phi}_{b\chi} \end{array} \right)^+\, \hat{p}^2 \, \left(\begin{array}{cc} N_c I(m_t,0, \hat{p}) & 0 \\ 0 & N_c I(m_\chi,0, \hat{p}) \end{array}\right)\left(\begin{array}{c}{\Phi}_{bt}\\ {\Phi}_{b\chi} \end{array} \right)\nonumber\\ && + \int d^4 x \left(\begin{array}{c}\Phi_{tt}\\ \Phi_{t\chi} \end{array} \right)^+\, \hat{p}^2 \, \left(\begin{array}{cc} N_c I(m_t,m_t, \hat{p})  & 0 \\ 0 & N_c I(m_t,m_\chi, \hat{p})  \end{array}\right)\left(\begin{array}{c}\Phi_{tt}\\ \Phi_{t\chi} \end{array} \right)\nonumber\\ && +
\int d^4 x \left(\begin{array}{c}\Phi_{\chi t}\\ \Phi_{\chi \chi} \end{array} \right)^+\, \hat{p}^2 \, \left(\begin{array}{cc} N_c I(m_t,m_\chi, \hat{p}) & 0 \\ 0 & N_c I(m_\chi,m_\chi, \hat{p})  \end{array}\right)\left(\begin{array}{c}\Phi_{\chi t}\\ \Phi_{\chi\chi} \end{array} \right) - {\cal V}(\hat{p},\Phi),\label{SEFFC}
\end{eqnarray}
\end{widetext}
where potential ${\cal V}(\hat{p},\Phi)$ depends on momentum operator as well as on the scalar fields. ${\cal V}(0,\Phi)\equiv {\cal V}(\Phi)$ has its minimum at $\langle \Phi_{tt} \rangle = \frac{v_t}{\sqrt{2}}=m_t$ and $\langle \Phi_{\chi\chi} \rangle = \frac{u_\chi}{\sqrt{2}}=m_\chi$. We are not interested in the particular form of $\cal V$.

In order to calculate the gauge boson masses we should substitute $\hat{p} \rightarrow \hat{p} - A$, where $A$ is the corresponding gauge field. In the tree level we should then substitute the scalar fields by the condensates, and omit $\hat{p}$. The mass term with the gauge field squared originates from the factor $\hat{p}^2$ of the above expression if the integrals $I(m_1,m_2,p)$ would be constants. Since these integrals are slow - varying logarithmic - like functions, for the  evaluation of the gauge boson masses we are able to substitute them by the values $I(m_1,m_2,\bar{p})$ for a certain typical value of momentum $\bar{p}$. For example, for $\Lambda = 1000$ TeV and $m_\chi = 17.5$ TeV (and for $\Lambda = 10 $ TeV and $m_\chi = 1.75$ Tev ) the difference between the values $N_c I(m_t,m_t,0)$, $N_c I(m_t,m_t, M_H)$, and $N_c I(m_t,m_t, i M_H)$ is within 1 per cent. The typical value of $\bar{p}^2$ in this problem is, in turn, of the order of the gauge boson mass squared, which is of the same order as $M_H^2$. Therefore, instead of $N_c I(m_a,m_b,p)$ in the following we substitute constants $Z^2_{abH}$.

The mass eigenstates $\chi_L$ and $t_L$ are composed of the original $\chi^{\prime}_L$ and $t^{\prime}_L$:
\begin{eqnarray}
\chi_L &=& - {\rm sin}\, \theta \, t^{\prime}_L +  {\rm cos}\, \theta \, \chi^{\prime}_L\nonumber\\
t_L &=&  {\rm }\,{\rm cos}\, \theta \, t^{\prime}_L +  {\rm sin}\, \theta \, \chi^{\prime}_L
\end{eqnarray}
These is the field $\left(\begin{array}{c}b^\prime_L\\t^{\prime}_L\end{array}\right)$, which carries the quantum numbers of the SM $SU(2)_L$ left - handed doublets. At the same time $t^\prime_R$, $\chi^\prime_L$, $\chi^\prime_R$ carry the quantum numbers of the right - handed top  quark.
Correspondingly, we represent
\begin{eqnarray}
\Phi_{\chi t} &=& - {\rm sin}\, \theta \, \Phi_{t^\prime_L t} +  {\rm cos}\, \theta \, \Phi_{\chi^{\prime}_L t}\nonumber\\
\Phi_{\chi \chi} &=& - {\rm sin}\, \theta \, \Phi_{t^\prime_L \chi} +  {\rm cos}\, \theta \, \Phi_{\chi^{\prime}_L \chi}\nonumber\\
\Phi_{t t}&=&  {\rm }\,{\rm cos}\, \theta \, \Phi_{t^{\prime}_L t} +  {\rm sin}\, \theta \, \Phi_{\chi^{\prime}_L t}
\nonumber\\
\Phi_{t \chi}&=&  {\rm }\,{\rm cos}\, \theta \, \Phi_{t^{\prime}_L \chi} +  {\rm sin}\, \theta \, \Phi_{\chi^{\prime}_L \chi}
\end{eqnarray}
This gives
\begin{widetext}
\begin{eqnarray}
S &\approx&\int d^4 x \left(\begin{array}{c}{\Phi}_{bt}\\ {\Phi}_{b\chi} \end{array} \right)^+\, \hat{p}^2 \, \left(\begin{array}{cc} Z_{t0 H}^2 & 0 \\ 0 & Z_{\chi 0 H}^2 \end{array}\right)\left(\begin{array}{c}{\Phi}_{bt}\\ {\Phi}_{b\chi} \end{array} \right)\nonumber\\ && +  \int d^4 x \left(\begin{array}{c}\Phi_{t^\prime_L t}\\ \Phi_{t^{\prime}_L \chi} \end{array} \right)^+\, \hat{p}^2 \, \left(\begin{array}{cc} Z_{t\chi H}^2 {\rm sin}^2\theta + Z_{tt H}^2 {\rm cos}^2 \theta & 0 \\ 0 & Z_{\chi\chi H}^2 {\rm sin}^2\theta + Z_{t\chi H}^2 {\rm cos}^2 \theta \end{array}\right)\left(\begin{array}{c}\Phi_{t^\prime_L t}\\ \Phi_{t^\prime_L \chi} \end{array} \right)\nonumber\\ && +\int d^4 x \left(\begin{array}{c}\Phi_{t^\prime_L t}\\ \Phi_{t^{\prime}_L \chi} \end{array} \right)^+\, \hat{p}^2 \, \left(\begin{array}{cc}  \frac{1}{2}{\rm sin}\,2\theta ( Z_{tt H}^2 - Z_{t\chi H}^2) & 0 \\ 0 &  \frac{1}{2}{\rm sin}\,2\theta ( Z_{t\chi H}^2 - Z_{\chi\chi H}^2)  \end{array}\right)\left(\begin{array}{c}\Phi_{\chi^\prime_L t}\\ \Phi_{\chi^\prime_L  \chi} \end{array} \right)
\nonumber\\ && +\int d^4 x \left(\begin{array}{c}\Phi_{\chi^\prime_L t}\\ \Phi_{\chi^\prime_L  \chi} \end{array} \right)^+\, \hat{p}^2 \, \left(\begin{array}{cc}  \frac{1}{2}{\rm sin}\,2\theta ( Z_{tt H}^2 - Z_{t\chi H}^2) & 0 \\ 0 & \frac{1}{2}{\rm sin}\,2\theta ( Z_{t\chi H}^2 - Z_{\chi\chi H}^2) \end{array}\right)\left(\begin{array}{c}\Phi_{t^\prime_L t}\\ \Phi_{t^{\prime}_L \chi} \end{array} \right)
\nonumber\\ &&+
\int d^4 x \left(\begin{array}{c}\Phi_{\chi^\prime_L t}\\ \Phi_{\chi^\prime_L \chi} \end{array} \right)^+\, \hat{p}^2 \, \left(\begin{array}{cc} Z_{tt H}^2 {\rm sin}^2\theta + Z_{t\chi H}^2 {\rm cos}^2 \theta & 0 \\ 0 & Z_{\chi\chi H}^2 {\rm cos}^2\theta + Z_{t\chi H}^2 {\rm sin}^2 \theta \end{array}\right)\left(\begin{array}{c}\Phi_{\chi^\prime_L t}\\ \Phi_{\chi^\prime_L  \chi} \end{array} \right)\nonumber\\ && - {\cal V}(\Phi),\label{SEFFC}
\end{eqnarray}
\end{widetext}

In this basis ($t_L^{\prime}, \chi_L^\prime, t_R, \chi_R$) the vacuum averages are:
\begin{widetext}
\begin{eqnarray}
\left(\begin{array}{cc}\langle \Phi_{t^\prime_L t}\rangle & \langle\Phi_{t^\prime_L \chi}\rangle \\ \langle \Phi_{\chi^\prime_L t}\rangle & \langle \Phi_{\chi^\prime_L \chi} \rangle
\end{array}\right) = \left(\begin{array}{cc}  \frac{1}{\sqrt{2}} v_t \, {\rm cos}\, \theta &- \frac{1}{\sqrt{2}} u_{\chi} \, {\rm sin}\, \theta\\
  \frac{1}{\sqrt{2}} v_t \, {\rm sin}\,\theta &  \frac{1}{\sqrt{2}} u_{\chi}\, {\rm cos}\,\theta
\end{array}\right)
\end{eqnarray}
\end{widetext}
The fields $\Phi_{t^\prime_L t}$ and $\Phi_{t^\prime_L \chi}$ are  transformed under the action of the SM gauge group while $\Phi_{\chi^\prime_L t}$ and $\Phi_{\chi^\prime_L \chi}$ are not.
In order to calculate the gauge boson masses induced by the scalar fields, we need to keep in the effective action the terms proportional to $p^2$ standing at the products of $\Phi^\prime_{t^\prime_Lt_R}$ and $\Phi^\prime_{t^\prime_L \chi_R}$:
\begin{eqnarray}
&&S_{p^2,t^\prime_L} =  \int {d^4x} \Phi^\prime_{t^\prime_L\chi_R} \hat{p}^2 (Z_{\chi\chi H}^2 {\rm sin}^2\theta + Z_{t\chi H}^2 {\rm cos}^2 \theta )\Phi^\prime_{t^\prime_L\chi_R}\label{sp2}\\&&+
\int {d^4x} \Phi^\prime_{t^\prime_Lt_R} \hat{p}^2 (Z_{t\chi H}^2 \,{\rm sin}^2\, \theta + Z_{tt H}^2 \,{\rm cos}^2\, \theta )\Phi^\prime_{t^\prime_Lt_R}\nonumber
\end{eqnarray}
In this expression we should substitute $\langle \Phi^\prime_{t^\prime_Lt_R} \rangle = v_t\, {\rm cos}\,\theta$ and $\langle \Phi^\prime_{t^\prime_L \chi_R}\rangle = - u_\chi\, {\rm sin}\, \theta$. At the same time we substitute $\hat{p}^2$ by the gauge field squared $A^2 = \frac{1}{4}(2g_W^2 W^+_\mu W^\mu + g^2_Z Z_\mu Z^\mu)$. Then Eq. (\ref{sp2}) gives the masses of W and Z bosons $M_Z = g_Z \eta/2$ and $M_W = g_W \eta/2$, where
\begin{eqnarray}
\eta^2 &=& v_t^2 {\rm cos}^2 \theta (Z_{tt H}^2 {\rm cos}^2 \theta + Z_{t\chi H}^2 {\rm sin}^2 \theta) \nonumber\\&&+ u^2_{\chi}\, {\rm sin}^2 \theta\, ( Z_{\chi\chi H}^2 {\rm sin}^2\theta + Z_{t\chi H}^2 {\rm cos}^2 \theta )   \nonumber\\ &\approx & 2 Z_{tt H}^2 m_t^2 \Big(1 + \frac{g^2_{t\chi}}{g^2_{\chi}}\frac{Z_{t\chi H}^2}{Z_{tt H}^2}\Big)\label{eta}
\end{eqnarray}
(We neglect here the terms proportional to $m^2_t/m^2_\chi$.) The $W$ and $Z$ - bosons acquire their observable masses if
$\eta \approx 246$ GeV.
In principle, this expression works reasonably well even for $\Lambda = 10$ TeV, $m_\chi = 10\, m_t$.

{\it Notice, that in our approach the two composite scalar fields $\Phi_{tt}$ and $\Phi_{\chi\chi}$ are condensed and both contribute to the gauge boson masses. While the condensate of $\Phi_{\chi\chi}$ (proportional to the mass of the heavy fermion $\chi$) is larger, that the condensate of $\Phi_{tt}$, the coupling of $\Phi_{\chi\chi}$ to the W and Z bosons is suppressed by the factor $m_t/m_\chi$. Thus, in general case the contributions of both scalars to the gauge boson masses are of the same order. For the large values of $\Lambda$ the $\Phi_{tt}$ dominates while for low values of $\Lambda$ the $\Phi_{\chi\chi}$ dominates. The $125$ GeV Higgs boson is composed mostly of $\Phi_{tt}$ and $\Phi_{t\chi}$. Therefore, for low scale of the hidden interaction its contribution to the Electroweak symmetry breaking is not dominant.   }

\subsubsection{Example choices of parameters}
\label{SectPart}

   Below we consider the two particular example choices of parameters, which give realistic spectrum of the scalar boson masses.

\begin{enumerate}

\item{}
Let us suppose first, that the scale of the new interaction is $\Lambda \sim 10^3$ TeV while $m_\chi = 100\, m_t$. We require
\begin{eqnarray}
M_H\approx m_{t}/\sqrt{{2}} \approx 125 \, {\rm GeV}\label{125}
\end{eqnarray}
and consider as an example the following particular choice of parameters (that provides Eqs. (\ref{eta}) and (\ref{125})):
\begin{eqnarray}
&& g_{t\chi} =  \, g_{\chi}\, \frac{Z_{tt H}}{Z_{t\chi H}}\, \sqrt{\frac{1}{Z_{ttH}^2}-1}, \label{cap2} \\ && g_{\chi} = 0.379\, Z_{t\chi H}^2 m_{\chi}^2,\quad g_t = 1.74\, Z_{t\chi H}^2 m_{\chi}^2\nonumber
\end{eqnarray}
All
values of bare and intermediate coupling constants as well as all observable masses for this example choice of initial parameters are collected in Table \ref{table}.

\item{}

The second example choice of parameters corresponds to $\Lambda = 10$ TeV and $m_\chi = 10 m_t$.
In this case we consider the following particular choice of parameters (that provides Eqs. (\ref{eta}) and (\ref{125})):
\begin{eqnarray}
&& g_{t\chi} =  \, g_{\chi}\, \frac{Z_{tt H}}{Z_{t\chi H}}\, \sqrt{\frac{1}{Z_{ttH}^2}-1}, \label{cap2} \\ && g_{\chi} = 0.169\, Z_{t\chi H}^2 m_{\chi}^2,\quad g_t = 1.74\, Z_{t\chi H}^2 m_{\chi}^2\nonumber
\end{eqnarray}
All
values of bare and intermediate coupling constants as well as all observable masses for this example choice of initial parameters are collected in Table \ref{table4}.

\end{enumerate}

Recall, that the values of $g_t$, $g_\chi$, $g_{t\chi}$ are the elements of matrix $G$ in the basis, in which the fermion mass matrix is diagonal. The original parameters of the model $g_{t,\chi,t\chi}^{(0)}$ are the elements of matrix $G$ in the basis, in which $(b^{\prime}_L t^{\prime}_L)^T$ is the $SU(2)_L$ doublet,  $\chi^{\prime}_L$ is the $SU(2)_L$ singlet while matrix $\Omega$ is diagonal. (Here $SU(2)_L$ is the part of the SM gauge group.) The values $g_{t,\chi,t\chi}^{(0)}$ are related to $g_{t,\chi,t\chi}$ via Eq. (\ref{G3A}) while $\alpha$ is given by Eq. (\ref{alpha}).
Parameters $\omega_{t,\chi}$ are related to the values of masses through gap equations Eq. (\ref{gapeqs}) and are of the order of $ \frac{N_c }{8 \pi^2}\, \Lambda^2$ that is much larger than the other quantities we encountered here. The original parameters are related to $\omega_{t,\chi}$ as  $\omega_{t,\chi} = {\rm cos}^2 \alpha \, \omega^{(0)}_{t,\chi} + {\rm sin}^2\alpha\, \omega^{(0)}_{\chi,t}$ and are also of the order of $\frac{N_c }{8 \pi^2}\, \Lambda^2$. This is the difference between $\omega_{t,\chi}$ and $\frac{N_c }{8 \pi^2}\, \Lambda^2$ that together with the values of $g_{t,\chi,t\chi}$ define the dynamical fermion masses.  The angle $\theta$  relates mass eigenstates $t_L,\chi_L$ with the original states $t^{\prime}_L, \chi_L^{\prime}$ (where $t^{\prime}_L$ is transformed under the action of the SM $SU(2)_L$ gauge group).

In the first one of the above examples the difference of scales between $\Lambda \sim 10^3$ TeV, $m_{\chi} \sim 17.5$ TeV and $m_t \sim 175$ GeV implies a kind of fine tuning. Such a difference may survive in the theory only if the values of coupling constants are close to their critical values at which the chiral symmetry breaking occurs. Moreover, to provide this we are to disregard the higher order $1/N_c$ corrections. The latter implies that the given NJL model should be defined with the counterterms that cancel the dangerous terms of the order of $\sim \Lambda^2$ coming in the next to leading $1/N_c$ corrections. (As it was mentioned in the introduction we imply this kind of the NJL model. For the discussion of this issue see also \cite{VolovikZubkovHiggs,torsionZ,Z2014_MGB} and references therein.) Notice that the results of \cite{Yamawaki} are valid under the same assumptions.

In general case the masses of the remaining CP - even scalar bosons are of the order of $m_{\chi}$ if $g_{\chi}\sim m_{\chi}^2$ and may be made sufficiently large by the appropriate choice of the ratio $m_t/m_{\chi}$. Correspondingly, they are able to decay into the pairs of fermions, which results in the imaginary part of their masses. The masses of CP - odd scalar bosons depend on the additional parameters $b_t, b_{\chi}, b_{t\chi}$. Those parameters should be chosen large enough in order to provide the stability of vacuum. We may choose their values in such a way, that the corresponding masses are also of the order of $m_{\chi}$. The mass of the charged scalar boson is given by Eq. (\ref{charged}) that is approximately equal to $M^{(2)}_{h_th_{\chi}}\approx M^{(2)}_{A_t A_{\chi}}$. In the considered examples the CP - even pseudo - Goldstone boson - the candidate for the role of the $125$ GeV Higgs is the only stable composite boson and is sufficiently lighter than the other composite scalar states. Due to mixing all neutral scalar bosons (except the $125$ GeV scalar) are able to decay into the pair $\bar{t}t$. We do not exclude, that some of the composite scalar bosons may become stable if the scale of the interaction is lower, than $10$ TeV while the heavy fermion mass is smaller, than $1.75$ TeV: this may occur if the masses of the scalar bosons are smaller than $2 m_t$ (for the neutral scalar bosons) and $m_t + m_b \approx m_t$ (for the charged scalar boson).

\subsubsection{The Effective lagrangian for the decays of the CP - even Pseudo - Goldstone  boson (neglecting the ratio $m_t/m_\chi$)}

As it will be seen below, the decay probabilities of the given scalar boson do not contradict the present experimental constraints. The H - boson production cross - sections and the decays of the Higgs bosons are typically described by the effective lagrangian of the following form:
\begin{widetext}
\bea
\label{eq:1}
L_{eff}  &=  &
  c_W {2 m_W^2  \over \eta}  H  \,   W_\mu^+ W_\mu^-  +     c_Z {m_Z^2 \over \eta} H  \,  Z_\mu  Z_\mu
 + c^{}_{g}  {\alpha_s \over 12 \pi \eta} H \, G_{\mu \nu}^a G_{\mu \nu}^a  +  c^{}_{\gamma} { \alpha \over \pi \eta} H \, A_{\mu \nu} A_{\mu \nu} .
\eea
\end{widetext}
Here $G_{\mu\nu}$ and $A_{\mu\nu}$ are the field strengths of gluon and photon fields. We do not consider here the masses of the fermions other than the top quark and $\chi$. Therefore, we omit in this lagrangian the terms responsible for the coresponding decays.
This effective lagrangian should be considered at the tree level only and describes the channels $H \rightarrow gg, \gamma \gamma, ZZ, WW,$. The fermions and $W$ bosons have been integrated out in the terms corresponding to the decays $H\rightarrow \gamma \gamma, gg$, and their effects are included in the effective  couplings $c_g$ and $c_\gamma$. In the SM we have $c_Z=c_W = 1$, while $c_{g}   \simeq 1.03\,, c_{\gamma} \approx  -0.81$ (see \cite{status}).

Below we evaluate the mentioned coupling constants in our model neglecting the ratio $m_t/m_\chi$. We will demonstrate, that the result is given by the SM values. The corrections to these values, therefore, depend on the ratio $m_t/m_\chi$ and are small provided that this ratio is small. The evaluation of these corrections is out of the scope of the present paper.

Let us define the neutral scalar field given by the sum of the condensate and the fluctuation $H$ around the condensate:
\begin{eqnarray}
&&\Phi_H  \approx  \sqrt{2}\frac{Z_{ttH} \Phi^\prime_{tt} - \omega \frac{Z_{t\chi H}^2}{Z_{ttH}} \Phi^\prime_{t\chi}}{\sqrt{1+
w^2 \frac{Z_{t\chi H}^2}{Z^2_t}}}\label{Hconst1} \\ & \sim &\sqrt{2} \frac{Z_{ttH} (\bar{t}_L t_R + \bar{t}_R t_L) -
\omega \frac{Z_{t\chi H}^2}{Z_{ttH}} (\bar{t}_L \chi_R + \bar{\chi}_R t_L)}{\sqrt{1+
w^2 \frac{Z_{t\chi H}^2}{Z^2_t}}}\nonumber
\end{eqnarray}
Vacuum average of this field is
\begin{eqnarray}
\langle\Phi_H \rangle& \approx & \frac{ Z_{ttH} v_t}{\sqrt{1+
w^2 \frac{Z_{t\chi H}^2}{Z^2_t}}}
\end{eqnarray}

We also define the neutral scalar fields
\begin{eqnarray}
\Phi_{h_t h_\chi} & \approx & \sqrt{2} \frac{\omega Z_{t\chi H} \Phi^\prime_{tt} + Z_{t\chi H}  \Phi^\prime_{t\chi}}{\sqrt{1+
w^2 \frac{Z_{t\chi H}^2}{Z^2_{ttH}}}}\nonumber\\
\Phi_{\varphi_t } & \approx & \sqrt{2} Z_{t\chi H} \Phi^\prime_{\chi t}
\nonumber\\
\Phi_{\varphi_\chi } & \approx & \sqrt{2} Z_{\chi \chi H} \Phi^\prime_{\chi \chi}
\label{Hconst2}
\end{eqnarray}
The latter field has vacuum average
\begin{eqnarray}
\langle\Phi_{\varphi_\chi} \rangle& \approx &  Z_{\chi \chi H} u_\chi
\end{eqnarray}

In order to calculate the decay constants of the Higgs boson we should substitute into  Eq. (\ref{sp2}) the following expressions
\begin{eqnarray}
\Phi_{t^\prime_L t}&=&  {\rm }\,{\rm cos}\, \theta \, \Phi_{t t} -  {\rm sin}\, \theta \, \Phi_{\chi t}
\nonumber\\
\Phi_{t^\prime_L \chi}&=&  {\rm }\,{\rm cos}\, \theta \, \Phi_{t \chi} -  {\rm sin}\, \theta \, \Phi_{\chi \chi}
\end{eqnarray}
This gives
\begin{widetext}
\begin{eqnarray}
&&S_{p^2,t^\prime_L} =  \int d^4x({\rm cos}\, \theta \, \Phi_{t \chi} -  {\rm sin}\, \theta \, \Phi_{\chi \chi}) \hat{p}^2 (Z_{\chi\chi H}^2 {\rm sin}^2\theta + Z_{t\chi H}^2 {\rm cos}^2 \theta )({\rm cos}\, \theta \, \Phi_{t \chi} -  {\rm sin}\, \theta \, \Phi_{\chi \chi})\label{sp3}\\&&+
\int d^4x({\rm cos}\, \theta \, \Phi_{t t} -  {\rm sin}\, \theta \, \Phi_{\chi t}) \hat{p}^2 (Z_{t\chi H}^2 \,{\rm sin}^2\, \theta + Z_{tt H}^2 \,{\rm cos}^2\, \theta )({\rm cos}\, \theta \, \Phi_{t t} -  {\rm sin}\, \theta \, \Phi_{\chi t})\nonumber
\end{eqnarray}
\end{widetext}

 The real parts of the scalar fields should be expressed through $\Phi_H$, $\Phi_{h_th\chi}$, $\Phi_{\varphi_t}$, and $\Phi_{\varphi_\chi}$:
 \begin{eqnarray}
\Phi^\prime_{t t}&=& \frac{\Big(\Phi_H + w \frac{Z_{t\chi H}}{Z_{ttH}} \Phi_{h_th_\chi} \Big)}{\sqrt{2}Z_{ttH}\sqrt{1+w^2\frac{Z_{t\chi H}^2}{Z_{ttH}^2}}}
\nonumber\\
\Phi^\prime_{t \chi}&=&  \frac{\Big(- \Phi_H w \frac{Z_{t\chi H}}{Z_{ttH}} + \Phi_{h_th_\chi} \Big)}{\sqrt{2}Z_{t\chi H}\sqrt{1+w^2\frac{Z_{t\chi H}^2}{Z_{ttH}^2}}}\nonumber\\
\Phi^\prime_{\chi t}  & \approx & \frac{\Phi_{\varphi_t }}{\sqrt{2} Z_{t\chi H}}
\nonumber\\
\Phi^\prime_{\chi \chi} & \approx & \frac{\Phi_{\varphi_\chi }}{\sqrt{2} Z_{\chi \chi H}}\nonumber
\end{eqnarray}
Next, we expand them around the condensates and keep only the terms linear in  $H$:
\begin{eqnarray}
&&S_{p^2,H} =  \int d^4 x \frac{{\rm cos}\, \theta \, H w \frac{Z_{t\chi H}}{Z_{ttH}} }{ Z_{t\chi H}\sqrt{1+w^2\frac{Z_{t\chi H}^2}{Z_{ttH}^2}}}  \hat{p}^2 (Z_{\chi\chi H}^2 {\rm sin}^2\theta + Z_{t\chi H}^2 {\rm cos}^2 \theta )\, {\rm sin}\, \theta \, u_\chi \label{sp4}\\&&+
\int d^4 x \,\frac{{\rm cos}\, \theta \, H }{Z_{ttH}\sqrt{1+w^2\frac{Z_{t\chi H}^2}{Z_{ttH}^2}}}  \hat{p}^2 (Z_{t\chi H}^2 \,{\rm sin}^2\, \theta + Z_{tt H}^2 \,{\rm cos}^2\, \theta )\, {\rm cos}\, \theta \, v_t \nonumber
\end{eqnarray}
Finally, we substitute $\hat{p}^2$ by the field $A^2 = \frac{1}{4}(2g_W^2 W^+_\mu W^\mu + g^2_Z Z_\mu Z^\mu)$:
\begin{widetext}
\begin{eqnarray}
S_{p^2,H} &=&  \int d^4 x \frac{H w^2 \frac{Z_{t\chi H}}{Z_{ttH}} }{ \sqrt{1+w^2\frac{Z_{t\chi H}^2}{Z_{ttH}^2}}} A^2  Z_{t\chi H}  \, v_t \label{sp4}\\&&+
\int d^4 x \,\frac{ H }{\sqrt{1+w^2\frac{Z_{t\chi H}^2}{Z_{ttH}^2}}}  A^2  Z_{tt H} \,  v_t \nonumber\\
 &\approx & \int d^4 x \, H v_t Z_{ttH} \sqrt{1+w^2\frac{Z_{t\chi H}^2}{Z_{ttH}^2}}\,A^2\nonumber\\
&\approx & \int d^4 x \, H \eta \,A^2
\end{eqnarray}
\end{widetext}
Recall that $M_Z = g_Z \eta/2$ and $M_W = g_W \eta/2$.
Thus we are able to evaluate the values of $c_W$ and $c_Z$ entering Eq. (\ref{eq:1}):
\begin{equation}
|c_W|^2 = |c_Z|^2 = 1 \label{ca}
\end{equation}
In order to evaluate constant $c_g$ we need to consider the vertex for the transition $H \rightarrow \bar{t}t$. It appears from the interaction term of the lagrangian
\begin{equation}
L_{\Phi\rightarrow \bar{t}t} = - \Bigl[
\bar{t}_L
\Phi_{tt}  t_R + (h.c.)\Bigr]
\end{equation}
This gives the interaction term of $H$ and the top - quark:
\begin{equation}
L_{H\rightarrow \bar{t}t} = - \frac{H}{\sqrt{2}Z_{ttH}\sqrt{1+w^2\frac{Z_{t\chi H}^2}{Z_{ttH}^2}}}\bar{t}{t}= - \frac{m_t}{\eta}\bar{t}{t}\,H
\end{equation}
and results in the Standard Model value
\begin{equation}
|c_g|^2=1
\end{equation}
Expression for $c_{\gamma}$ is more complicated. However, in the considered approximation (when we neglect corrections proportional to $m^2_t/m_\chi^2)$) it is also given by the SM value. Notice, that the top quark is integrated out in Eq. (\ref{eq:1}), and its coupling to $H$ is absorbed by $c_g$ and $c_{\gamma}$.

In principle, if we consider the choice of coupling constants that corresponds to sufficiently light $\chi$, the valuable corrections to the Higgs boson decay constants would appear. The corresponding experimental data are presented in Fig. 25 of \cite{CMStest}.

{\it Thus we see, that although the contribution of the $125$ GeV Higgs to the Electroweak symmetry breaking may not be dominant, its decay constants are close to their values in the Standard Model, where it gives the only contribtion to the gauge boson masses.}

{ It is worth mentioning, that in our estimates we disregarded completely the running of coupling constants from the scale $\Lambda$ to the electroweak scale. This running affects essentially the values of the scalar boson masses if the scale is sufficiently high  \cite{top,top2}. This is more or less obvious, however, that our large number of free parameters allows a choice that leads to the necessary relation between the renormalized values of scalar boson masses and renormalized values of effective coupling constants entering Eq. (\ref{eq:1}). }

We did not consider in this paper the other contributions of the Electroweak gauge fields to the effective lagrangian. Those contributions are suppressed, however, due to the smallness of the electroweak gauge coupling (see \cite{dobrescu,Yamawaki}). We also did not considered the contribution of the heavy fermion $\chi$ to the Electroweak polarization operators (S and T parameters). The latter contribution is controlled by the ratio $m_t/m_{\chi}$ and if its value is sufficiently small the contribution of $\chi$ to S and T parameters is suppressed \cite{Yamawaki}.

\begin{widetext}
\begin{table}
\begin{center}
\begin{small}
Bare parameters\\
\begin{tabular}{|c|c|c|c|c|c|c|c|c|c|c|c|c|c|c|c|c|}
\hline
$\omega_t^{(0)}-\frac{N_c}{8 \pi^2}\, \Lambda^2$ & $\omega^{(0)}_\chi-\frac{N_c}{8 \pi^2}\, \Lambda^2$ & $g^{(0)}_t$ & $g^{(0)}_{t\chi}$ & $g^{(0)}_\chi$ & $b^{(0)}_t$ & $b^{(0)}_{t\chi}$ & $b^{(0)}_\chi$ & $\Lambda$  \\
\hline
$87$ TeV $^2$ & $- 84$ TeV $^2$ & $106$ TeV $^2$ & $18$ TeV $^2$ &$5.9$ TeV $^2$ & $563$ TeV $^2$ & $33$ TeV $^2$ &$-0.073$ TeV $^2$ &$1000$ TeV \\
\hline
\end{tabular}\\\vspace{3mm}
Intermediate parameters\\
\begin{tabular}{|c|c|c|c|c|c|c|c|c|c|c|c|c|c|c|c|c|}
\hline
$\omega_t^{}-\frac{N_c}{8 \pi^2}\, \Lambda^2$ & $\omega^{}_\chi -\frac{N_c}{8 \pi^2}\, \Lambda^2$& $g^{}_t$ & $g^{}_{t\chi}$ & $g^{}_\chi$ & $b^{}_t$ & $b^{}_{t\chi}$ & $b^{}_\chi$ & $f_t$ & $f_\chi$  \\
\hline
$78$ TeV $^2$ & $- 74$ TeV $^2$ & $92$ TeV $^2$ & $39$ TeV $^2$ &$20$ TeV $^2$ & $528$ TeV $^2$ & $105$ TeV $^2$ &$264$ TeV $^2$  & $77$ TeV $^2$ & $20$ TeV $^2$   \\
\hline
\end{tabular}
\\ \vspace{3mm}
Fermion masses, scalar boson masses,  and mixing angles\\
\begin{tabular}{|c|c|c|c|c|c|c|c|c|c|c|c|c|c|c|c|c|}
\hline
$m_t^{}$ & $m_\chi$  &$M_H$ &  $M^{(2)}_{h_t h_{\chi}}$ & $ M^{(2)}_{A_t A_{\chi}}$ & $M^{(1)}_{H^{\pm}_t, H^{\pm}_{\chi}}$  \\
\hline
$175$ GeV & $17.5$ TeV  & $125$ GeV & $(22-2.9\,i)$ TeV & $(22-2.9\,i)$ TeV  & $(22-2.9i)$ TeV  \\
\hline
$M^{}_{\varphi_t}$ & $M^{}_{\varphi_{\chi}}$ & $M^{(1)}_{\pi_t,\pi_{\chi}}$ & $ M^{(2)}_{\pi_{\chi}, \pi_t}$ & $\alpha$ & $\theta$ \\
\hline
$(22-0.5\,i)$ TeV & $35$ TeV  & $(63-10\,i)$ TeV  &$(38-7\,i)$ TeV  & $-0.0763\, \pi$ & $0.00627\,\pi$ \\
\hline
\end{tabular}
\end{small}
\caption{Values of bare and intermediate coupling constants as well as the observable masses for the first considered example choice of initial parameters. Bare coupling constants enter the original lagrangian: Eqs. (\ref{LI0}), (\ref{Omega0}), (\ref{gtc3}), (\ref{gtc4}). The ultraviolet cutoff $\Lambda$ is present there implicitly. Intermediate coupling constants appear, when the lagrangian is written in terms of mass eigenstates. Those parameters enter gap equation Eq. (\ref{gapeqs}) and the expressions for scalar boson masses. Mixing angles $\alpha$ and $\theta$ enter the relation between the original fermion fields of the model and the mass eigenstates in Eqs. (\ref{rot}), (\ref{rot2}).
Accuracy of our calculations is within about 5 per cents for the considered choice of parameters. All scalar bosons excluding the $125$ GeV Higgs are unstable, which corresponds to their decay into the pairs of fermions. Correspondingly, their masses have imaginary parts. The imaginary part of $M_{\varphi_\chi}$ is suppressed by the factor $m_t/m_\chi$ and is not represented here.   } \label{table}
\end{center}
\end{table}
\end{widetext}

\begin{widetext}
\begin{table}
\begin{center}
\begin{small}
Bare parameters\\
\begin{tabular}{|c|c|c|c|c|c|c|c|c|c|c|c|c|c|c|c|c|}
\hline
$\omega_t^{(0)}-\frac{N_c}{8 \pi^2}\, \Lambda^2$ & $\omega^{(0)}_\chi-\frac{N_c}{8 \pi^2}\, \Lambda^2$ & $g^{(0)}_t$ & $g^{(0)}_{t\chi}$ & $g^{(0)}_\chi$ & $b^{(0)}_t$ & $b^{(0)}_{t\chi}$ & $b^{(0)}_\chi$ & $\Lambda$  \\
\hline
$0.45$ TeV $^2$ & $- 0.38$ TeV $^2$ & $0.48$ TeV $^2$ & $0.063$ TeV $^2$ &$0.0094$ TeV $^2$ & $2.7$ TeV $^2$ & $0.27$ TeV $^2$ &$-0.056$ TeV $^2$ &$10$ TeV \\
\hline
\end{tabular}\\\vspace{3mm}
Intermediate parameters\\
\begin{tabular}{|c|c|c|c|c|c|c|c|c|c|c|c|c|c|c|c|c|}
\hline
$\omega_t^{}-\frac{N_c}{8 \pi^2}\, \Lambda^2$ & $\omega^{}_\chi -\frac{N_c}{8 \pi^2}\, \Lambda^2$& $g^{}_t$ & $g^{}_{t\chi}$ & $g^{}_\chi$ & $b^{}_t$ & $b^{}_{t\chi}$ & $b^{}_\chi$ & $f_t$ & $f_\chi$  \\
\hline
$0.43$ TeV $^2$ & $- 0.36$ TeV $^2$ & $0.45$ TeV $^2$ & $0.14$ TeV $^2$ &$0.044$ TeV $^2$ & $2.6$ TeV $^2$ & $0.5$ TeV $^2$ &$1.3$ TeV $^2$  & $0.44$ TeV $^2$ & $0.044$ TeV $^2$   \\
\hline
\end{tabular}
\\ \vspace{3mm}
Fermion masses, scalar boson masses,  and mixing angles\\
\begin{tabular}{|c|c|c|c|c|c|c|c|c|c|c|c|c|c|c|c|c|}
\hline
$m_t^{}$ & $m_\chi$  &$M_H$ &  $M^{(2)}_{h_t h_{\chi}}$ & $ M^{(2)}_{A_t A_{\chi}}$ & $M^{(1)}_{H^{\pm}_t, H^{\pm}_{\chi}}$  \\
\hline
$175$ GeV & $1.75$ TeV  & $125$ GeV & $(2.0-0.5\,i)$ TeV & $(2.0-0.5\,i)$ TeV  & $(2.0-0.5\, i)$ TeV  \\
\hline
$M^{}_{\varphi_t}$ & $M^{}_{\varphi_{\chi}}$ & $M^{(1)}_{\pi_t,\pi_{\chi}}$ & $ M^{(2)}_{\pi_{\chi}, \pi_t}$ & $\alpha$ & $\theta$ \\
\hline
$(2.3-0.1\,i)$ TeV & $3.5$ TeV  & $(5.8-2\,i)$ TeV  &$(3.5-1\,i)$ TeV  & $-0.054\, \pi$ & $0.0098\,\pi$ \\
\hline
\end{tabular}
\end{small}
\caption{Values of bare and intermediate coupling constants as well as the observable masses for the second considered example choice of initial parameters. Bare coupling constants enter the original lagrangian: Eqs. (\ref{LI0}), (\ref{Omega0}), (\ref{gtc3}), (\ref{gtc4}). The ultraviolet cutoff $\Lambda$ is present there implicitly. Intermediate coupling constants appear, when the lagrangian is written in terms of mass eigenstates. Those parameters enter gap equation Eq. (\ref{gapeqs}) and the expressions for scalar boson masses. Mixing angles $\alpha$ and $\theta$ enter the relation between the original fermion fields of the model and the mass eigenstates in Eqs. (\ref{rot}), (\ref{rot2}).
Accuracy of our calculations is within about 15 per cents for the considered choice of parameters. All scalar bosons excluding the $125$ GeV Higgs are unstable, which corresponds to their decay into the pairs of fermions. Correspondingly, their masses have imaginary parts. The imaginary part of $M_{\varphi_\chi}$ is suppressed by the factor $m_t/m_\chi$ and is not represented here.   } \label{table4}
\end{center}
\end{table}
\end{widetext}

\section{Conclusion and discussions}

\label{sectconclusions}

In the considered scenario, the symmetry breaking takes place at the high energy scale,
where there is the hidden symmetry (in  $^3$He-B it is the separation of
spin and orbital rotations, in the proposed model of top quark condensation this is the $SU(3)_L$ symmetry). This symmetry is violated at low energy. As a
result, some of the Nambu-Goldstone modes transform to the light Higgs
bosons.
Such scenarios of emergence of light Higgs may have some, though not always
exact, parallels in the other models of high energy physics.

Let us consider, for example, the hidden chiral symmetry in QCD. It is provided by
an approximation in which the $u$ and $d$ quarks are
considered as massless. The spontaneous breaking of the hidden symmetry
leads to three pions (one neutral and two charged) as the massless Goldstone
bosons. These pions become massive when one takes into
account the nonzero masses of $u$ and $d$ quarks. The masses of pions
are  much smaller, than the mass of the local Higgs boson (the
$\sigma$-meson). This situation is similar to that of the top - seesaw models of \cite{dobrescu,Yamawaki}, where the explicit mass term is introduced that breaks the hidden $SU(3)_L$ symmetry. It, however, is different from that of
$^3$He-B, where there is no explicit mass term for the fermions. Instead, the spin - orbit interaction appears as a modification of the original four - fermion interaction. In the present paper we propose the model of top quark condensation, in which the $SU(3)_L$ symmetry is broken by the modification of the four - fermion interaction in an analogy with $^3$He-B.

The top quark condensation model considered in the present paper is similar to the top - seesaw models of \cite{dobrescu,Yamawaki}. Our model (as well as the models of \cite{dobrescu,Yamawaki}) contains the CP - even light Higgs, whose mass appears as a result of the soft breakdown of $SU(3)_L$ symmetry. In this respect this model differs from QCD, where the massive pions are CP - odd states.  The light Higgs of our model is similar to the light Higgs boson of $^3$He-B, that has all the signatures of the Higgs
boson:  it is the amplitude mode of the Higgs triplet vector  field ${\bf
n}$, while the rotational modes of Higgs triplet represent the NG bosons in
a full correspondence with the Higgs scenario.

The situation in  $^3$He-B and in the complicated top quark condensation model considered here is also close to that of the Little Higgs models
 (see review \cite{LittleHiggsReview2005} and
references therein). In the Little Higgs approach the Higgs particles also
appear as the pseudo-NG bosons (though not composed of the top quark). The corresponding field has all the
properties of the Higgs field, whose collective modes contain both the
amplitude Higgs modes (the Higgs bosons) and the NG modes (in gauge
theories the NG modes are absorbed by the gauge fields and become the
massive gauge bosons). That is why we may also say that the massive mode $\#$15 in  $^3$He-B --
the gapped spin wave -- represents the condensed matter analog of the
Little Higgs. The appearance of the analogs of the Little Higgs bosons is also possible  in the other condensed
matter systems. The abstracts of the recent International Workshop "Higgs
Modes in Condensed Matter and Quantum Gases", can be found in Ref.
\cite{abstractsKyoto2014}.

In $^3$He-B, there is the large difference in energy scales between the
heavy Higgs bosons and the light Little Higgs. That is why the
transformation of the NG mode to the Little Higgs practically does not
violate the Nambu sum rule \cite{Nambu1985}. The Nambu partner of the Little
Higgs is the heavy Higgs with energy close to $2\Delta$, which has the same
quantum numbers $(J=1,J_z=0)$, but different parity.
The considered light Higgs is essentially lighter than the fermionic
quasiparticles, which have the gap $\Delta$. This indicates, that if this scenario works in the SM
and the observed 125 GeV Higgs is the Pseudo - Goldstone boson, then there should be the additional fermion,
 which is much heavier, than the top quark.

Indeed, in the considered model of top quark condensation the additional fermion $\chi$ is much more heavy than the top quark. In the proposed model we evaluate in the leading order of the $1/N_c$ expansion the decay branching ratios of the Higgs boson. Their deviations from the SM values are suppressed by the ratios $m_t/m_\chi$, and therefore do not contradict the present LHC data. The CP even neutral pseudo - Goldstone boson may be composed mostly of the $\bar{t}_L t_R $ and $\bar{t}_L \chi_R$ pairs (with the valuable contribution of the first pair). The corresponding coupling constants in the effective lagrangian (that describe its decays) may be very close to the SM values. The parameters of the model may be chosen in such a way, that the Higgs boson mass is given by the observable value $125$ GeV.
In the present paper we do not analyse in details the phenomenology of the model. In particular, we do not consider the effect of the SM gauge interactions on the model and the mechanism for the generation of the masses of the other SM fermions. (Only the mechanism for the generation of $m_t$ has been discussed.) Besides, we disregarded completely the running of coupling constants from the scale $\Lambda$ to the electroweak scale. This running may affect essentially the values of the scalar boson masses if the scale $\Lambda$ is sufficiently high \cite{top,top2}. This is more or less obvious, however, that even in such case our large number of free parameters allows a choice that leads to the necessary relation between the renormalized values of scalar boson masses and renormalized values of effective coupling constants entering Eq. (\ref{eq:1}).
 On the other hand for low values of $\Lambda$ our estimate for the Higgs boson mass Eq. (\ref{mha}) becomes less accurate. Say, at $\Lambda = 10$ TeV and $m_\chi = 1.75$ TeV it gives accuracy about $10$ percent. However, the proposed approach clearly remains at work for $\Lambda$ equal to a few TeV. The detailed consideration of this case is technically rather complicated if we need to achieve a better accuracy of the estimates.
 Thus we expect, that our consideration may give a sufficient qualitative pattern of the theory, in which the pseudo - Goldstone boson plays the role of the $125$ GeV Higgs. We prefer not to call our construction the top - seesaw model because unlike \cite{topseesaw} the traditional scheme with the off - diagonal condensate $\langle \bar{t}_L \chi_R\rangle$ is not necessary (though allowed).

Unlike \cite{dobrescu,Yamawaki} in our case the explicit mass term is absent and the soft breaking of the $SU(3)$ symmetry is given solely by the four - fermion terms. This reveals the complete analogy with $^3$He, where there is no explicit mass term and the spin - orbit interaction has the form of the modification of the original four - fermion interaction.

The top quark condensation model with the four - fermion interaction considered here should necessarily appear as the effective low energy approximation to the unknown microscopic theory. Certain non - NJL corrections to various physical quantities are to appear from this microscopic theory. If the discussed scenario (in which the $125$ GeV Higgs boson appears as the composite Pseudo - Goldstone boson), will be confirmed by experiment, such a theory is to be constructed.
It may be very unusual. In particular, the nature of the forces binding fermions in Higgs boson may be related to such complicated objects as the emergent bosonic fields that exist within the fermionic condensed matter systems (graphene and superfluid He-3). In condensed matter systems various emergent gauge and gravitational fields appear \cite{emergent}. Those emergent gravitational fields should not be confused with the real gravitational fields. Typically, the emergent gravity in condensed matter does not have the main symmetry of the gravitational theory (the invariance under the diffeomorphisms does not arise). That's why in the majority of cases we may speak of the emergent gravity only as of the geometry experienced by the fermionic quasiparticles. The fluctuations of the gravitational fields themselves are not governed by the diffeomorphism - invariant theory. We suppose that the objects like these emergent gauge and gravitational fields may play a certain role in the formation of forces binding fermions within the composite Higgs bosons.

We also do not exclude the possibility, that certain part of the extended real gravitational fields may play a role in the formation of such forces. In particular, there exist the theories of quantum gravity with torsion \cite{torsionZ}, in which the fluctuations of torsion have the scale slightly above 1 TeV while the scale of the fluctuations of metric is the Plank mass. The mentioned fluctuations of torsion may also be related to the formation of composite Higgs bosons.

A less unusual scenario of physics behind the four - fermion interactions of the top - seesaw model involves the exchange by massive gauge bosons, which appear in the conventional renormalizable field theory (see, for example, \cite{Z2014_MGB} and references therein).

It is worth mentioning, that our model, in principle, admits a generalization to the case, when all remaining SM quarks and leptons are present. In the framework of top - seesaw models the corresponding generalization has been discussed, for example, in  \cite{topseesaw}. In our case we should start from the generalization of Eqs. (\ref{LI0}) and (\ref{btchi}), where all left - handed and right - handed quarks and leptons are present. In addition the lagrangian may include several extra fermions $\chi^{(i)}$, $i = 1,2,...$ (similar to the $\chi$ of the present paper). The lagrangian should be invariant under the unitary transformation group $G$ that mixes left - handed quarks and leptons and the extra fields $\chi^{(i)}_L$. At the next step of the construction we should break this $G$ softly by the four - fermion interactions and, possibly, by the explicit mass terms that involve the extra fermions $\chi^{(i)}$. This will result in the appearance of the Pseudo - Goldstone bosons. The whole construction should provide the appearance of the CP - even Pseudo - Goldstone boson that may be identified with the $125$ GeV Higgs boson, while the remaining scalar bosons should have much larger masses (or much smaller production cross sections) in order to avoid the present experimental exclusions. From the technical point of view such a construction should be rather complicated.

M.A.Z. kindly acknowledges useful discussions with V.A.Miransky and the support of the University of Western Ontario, where this work was initiated. The work of M.A.Z. was  supported by Far Eastern Federal University, by Ministry of science and education of Russian Federation under the contract 02.A03.21.0003, and by grant RFBR 14-02-01261.
The work of GEV is supported by the Academy of Finland
through its LTQ CoE grant (Project No. 250280).

\end{document}